\documentclass[fleqn,usenatbib]{mnras}
\usepackage{graphicx}
\usepackage{amssymb}
\usepackage{amsmath}
\usepackage{mathabx}
\usepackage{txfonts}
\usepackage[T1]{fontenc}
\usepackage{ae,aecompl}
\usepackage{cleveref}

\usepackage[dvipsnames]{xcolor}

\title{Eccentricity driving of pebble accreting low-mass planets}

\author[D. A. Velasco Romero et al]{%
David A. Velasco Romero$^{1}$\thanks{E-mail: david.velasco@icf.unam.mx} , Fr\'ed\'eric S. Masset$^{2,3}$ , Romain Teyssier$^{1,4}$\\
$^{1}$Institute for Computational Science, University of Zurich, Winterthurerstrasse 190, 8057 Zurich, Switzerland\\
$^{2}$Instituto de Ciencias F\'isicas, Universidad Nacional Aut\'onoma de M\'exico, Av. Universidad s/n, 62210 Cuernavaca, Mor., Mexico\\
$^{3}$University Nice-Sophia Antipolis, CNRS, Observatoire de la C\^ote d'Azur, Laboratoire LAGRANGE, CS 34229. 06304 Nice Cedex 4, France\\
$^{4}$Department of Astrophysical Sciences, Princeton University,
4 Ivy Lane,
 Princeton, New Jersey 08544, United States.
}

\date{Accepted XXX. Received YYY; in original form ZZZ}

\pubyear{2021}

\begin{document}
\label{firstpage}
\pagerange{\pageref{firstpage}--\pageref{lastpage}}
\maketitle

\begin{abstract}
By means of high resolution hydrodynamical, three-dimensional calculations with nested-meshes, we evaluate the eccentricity reached by a low-mass, luminous planet embedded in an inviscid disc with constant thermal diffusivity and subjected to thermal forces. We find that a cell size of at most one tenth of the size of the region heated by the planet is required to get converged results. When the planet's luminosity is supercritical, we find that it reaches an eccentricity of order $10^{-2}$--$10^{-1}$, which increases with the luminosity and broadly scales with the disc's aspect ratio.
Restricting our study to the case of pebble accretion, we incorporate to our model the dependence of the accretion rate of pebbles on the eccentricity. 
There is therefore a feedback between eccentricity, which determines the accretion rate and hence the planet's luminosity, and the luminosity, which yields the eccentricity attained through thermal forces. We solve for the steady state eccentricity and study how this quantity depends on the disc's turbulence strength parameter $\alpha_z$, on the dimensionless stopping time of the pebbles $\tau_s$, on the inward mass flux of pebbles and on the headwind (the difference between the gas velocity and the Keplerian velocity). We find that in general low-mass planets (up to a few Earth masses) reach eccentricities comparable to the disc's aspect ratio, or a sizeable fraction of the latter. Eccentric, low-mass protoplanets should therefore be the norm rather than the exception, even if they orbit far from other planets or from large scale disturbances in the disc.
\end{abstract}


\defcitealias{2017MNRAS.465.3175M}{MV17}
\defcitealias{2019MNRAS.483.4383V}{paper~I}

\section{Introduction}
\label{sec:introduction}
The gravitational interaction between a forming planet and its parent protoplanetary disc is a key ingredient to account for the orbital evolution of the planet as it grows in the disc. The force exerted by the disc depends on the perturbation of the density field in the disc regions close to the planet. It has recently been shown that heat diffusion on the one hand, and the release of heat by the planet into the surrounding disc on the other hand, can considerably alter the density perturbation in the vicinity of low-mass planets, and as a consequence the force exerted by the disc onto the planet. With respect to the abundantly studied case of a planet that does not release heat and is embedded in a disc that behaves adiabatically (i.e. in which heat diffusion is negligible), new contributions arise as a consequence of these two new ingredients. These contributions are generically referred to as thermal forces or thermal torques. While they have important consequences even in the case of a planet on a circular orbit \citep{2014MNRAS.440..683L,2015Natur.520...63B,2017MNRAS.472.4204M} as they can modify considerably the speed and even the direction of migration \citep{2019MNRAS.486.5690G,2021MNRAS.501...24C}, they play an even more important role on the evolution of the eccentricity and inclination \citep{2017A&A...606A.114C, 2017arXiv170401931E}. \citet{2017arXiv170401931E} have found that non-luminous planets embedded in radiative discs (hence subjected to some form of heat diffusion) have their eccentricity and inclination damped much more vigorously than if the disc is adiabatic. However, they also found that if the planets are luminous and heat the surrounding disc at a rate given by the accretion of infalling, solid bodies, then the eccentricity and inclination are no longer damped but grow with time. Quite remarkably, the growth of eccentricity is largely faster than that of inclination, and the latter ceases once the eccentricity levels off to a finite, asymptotic value. The stationary state thus reached by a low-mass planet is that of a planet on an eccentric orbit, with a negligible inclination and an eccentricity comparable to the aspect ratio of the disc. In what follows, we will therefore neglect the inclination of the planet's orbit and will focus exclusively on the time evolution of its eccentricity. While the work of \citet{2017arXiv170401931E} gave a qualitative understanding of the mechanism responsible for the eccentricity growth, it was based on numerical simulations that barely resolved the region heated by the planet, and as a consequence misrepresented (and probably underestimated) the thermal force exerted by the disc. \citet{2019MNRAS.485.5035F} performed an analytic study of the growth of eccentricity arising from thermal forces, but this study is valid in the limit of an epicyclic excursion much smaller than the size of the heated region. This is clearly not the case when the planet reaches its asymptotic eccentricity, since the epicyclic excursion is of the order of the disc's pressure scalelength, itself generally much larger than the size of the heated region \citep{2017MNRAS.472.4204M,2019MNRAS.485.5035F}. For these reasons, it is clearly necessary to perform numerical simulations at much larger resolution than those of \citet{2017arXiv170401931E} to get converged results for the asymptotic eccentricity of a planet with a given luminosity. This is, in part, the purpose of the present work. In order to reach converged results, we had to resort to a nested-mesh setup. We then study how the eccentricity attained depends on the planet's mass, luminosity and on the disc's aspect ratio.

The luminosity of the planet itself comes from the heat released by infalling, solid material onto the planet. This material can either be under the form of pebbles (solid particles from millimetre sizes to at most a few metres in size) or planetesimals (solid bodies with much larger sizes, ranging from tens of metres to hundreds of kilometres). In this work we do not consider the accretion of planetesimals and restrict ourselves to that of pebbles, thought to be the principal mode of growth for low mass planets over most of the planet-forming regions of the disc \citep{2010A&A...520A..43O,2012A&A...544A..32L}. It has recently been established \citep{2018A&A...615A.138L} that the rate of pebble accretion onto a low-mass planet depends on its eccentricity. The accretion rate usually increases with the eccentricity up to values which are broadly in the range $[10^{-2},10^{-1}]$, as the planet sweeps up more pebbles, then decays for larger eccentricities, as the encounters between the planet and pebbles are then too short for those to efficiently settle toward the planet. We see therefore that there is a feedback loop between the eccentricity, that regulates the accretion rate and therefore the luminosity, and the luminosity, which determines the eccentricity via the thermal forces. We study this feedback in the second part of the present work, and determine the eccentricities and luminosities typically attained by low-mass planets in a variety of disc models, for different pebble sizes. We also assess how accretion is shortened with respect to the case of planets remaining on circular orbits.

This paper is organised as follows: in section~\ref{sec:description-problem} we introduce our notation and the governing equations of the hydrodynamical calculations. In section~\ref{sec:results-from-linear}, we give a brief summary of the results from linear theory about the eccentricity and inclination evolution of a luminous, low-mass planet.  We then provide some details about our numerical implementation in section~\ref{sec:numer-impl}, in particular our strategy for the layout of the nested meshes, and the parameters of our fiducial disc model.
In section~\ref{sec:results} we present the results of our hydrodynamical calculations. The main output on which we focus in this section is the eccentricity attained at larger time by a planet with given mass and luminosity. In section~\ref{sec:feed-back-eccentr}, we incorporate to our results the feedback of the eccentricity on the luminosity, through the accretion rate. Finally, in section~\ref{sec:disc-concl}, we summarise our results, enumerate the main shortcomings of our analysis, and present our conclusions.

\section{Description of the problem}
\label{sec:description-problem}

The system under study consists of a planet of mass $M_p$ and luminosity $L$ immersed in a protoplanetary disc of gas around a central star of mass $M_\star$. We consider the planet to be on an eccentric orbit with semi-major axis $a$ and eccentricity $e$. Our initial conditions are such that the gaseous disc has a surface density $\Sigma$ given by:
\begin{equation}
  \label{eq:5.1}
  \Sigma=\Sigma_0\left(\frac{r}{a}\right)^{-\alpha}
\end{equation} 
and a temperature $T$ given by:
\begin{equation}
  \label{eq:5.2}
  T=T_0\left(\frac{r}{a}\right)^{-\beta}
\end{equation}
The disc has an aspect ratio $h=H/r$, where $r$ is the distance to the central star and $H$ is the pressure scale-length, given by:
\begin{equation}
  \label{eq:6}
  H=\frac{c_s^\mathrm{iso}}{\Omega},
\end{equation}
where $\Omega$ is the local orbital frequency and $c_s^\mathrm{iso}$ is the local isothermal sound speed:
\begin{equation}
  \label{eq:7}
  c_s^\mathrm{iso}=\sqrt{\frac{{\cal R}T}{\mu}},
\end{equation}
with ${\cal R}$ being the ideal gas constant and $\mu$ the mean molecular mass.
The thermal diffusivity of the disc is denoted with $\chi$.
In the particular case of a vanishing eccentricity,
the planet's distance to corotation is:
\begin{equation}
   x^0_p = \eta h^2 a,
\end{equation}
where $\eta = (2\alpha+\beta+3)/6$.

\subsection{Governing equations}
\label{sec:governing-equations}
The governing equations of our system consist of the continuity equation, the equations of conservation of momentum and the equation on the density of internal energy. The continuity equation reads: 
\begin{equation}
    \label{eq:1}
    \partial_t \rho + \nabla \cdot (\rho\boldsymbol{V}) = 0,
  \end{equation}
  where $\rho$ is the mass density and $\boldsymbol{V}$ is the velocity vector. The equations of conservation of momentum read in a spherical system of coordinates $(r,\phi,\theta)$ rotating about its $z$-axis at the angular frequency $\Omega_f$:
  \begin{equation}
  \label{eq:8}
  \partial_t(\rho v_r) + \nabla \cdot (\rho  v_r\boldsymbol{V})=\rho\frac{(v_\phi+r\sin\theta\Omega_f)^2+
    v_\theta^2}{r}-\partial_rp-\rho\partial_r\Phi,
\end{equation}
\begin{equation}
  \label{eq:9}
  \partial_t(\rho j)+\nabla\cdot (\rho j\boldsymbol{V})=-\partial_\phi p-\rho\partial_\phi\Phi
\end{equation}
and
\begin{equation}
  \label{eq:11}
  \partial_t(\rho r v_\theta)+\nabla\cdot(\rho r v_\theta\boldsymbol{V})=\rho(v_\phi+r\sin\theta\Omega_f)^2\cot\theta-\partial_\theta p-\rho\partial_\theta\Phi,
\end{equation}
where $p$ is the pressure, $\Phi$ the gravitational potential, $v_r$, $v_\phi$ and $v_\theta$ the components of the velocity and $j=r\sin\theta v_\phi+r^2\sin^2\theta\Omega_f$ the specific angular momentum (as seen in the inertial frame).
Lastly, the equation for the internal energy density $\mathfrak{e}$ reads:
\begin{equation}
    \label{eq:4}
\partial_t \mathfrak{e} + \nabla \cdot (\mathfrak{e}\boldsymbol{V}) = -p\nabla \cdot \boldsymbol{V} - \nabla \cdot \left( \chi\rho\nabla \frac{\mathfrak{e}}{\rho} \right) + L \delta (\boldsymbol{r}-\boldsymbol{r_p}),
\end{equation}
where the second term of the right hand side is the divergence of the  heat flux given by Fourier's law, and the last term represents the energy released by the planet into the disc. We consider the disc to be inviscid, hence we do not consider a source term on the right hand side of Eq.~\eqref{eq:4} other than that of the planet. The system is closed by the equation of state of ideal gases:
\begin{equation}
  \label{eq:12}
p=(\gamma-1)\mathfrak{e},
\end{equation}
where $\gamma$ is the adiabatic index. Note that we have used, in all our hydrodynamical simulations, an inviscid disc. Should we have included some finite, physical viscosity, cooling prescriptions should also have been taken into account in order to avoid the accumulation of heat in the disc. The impact of a finite viscosity on the processes reported here should be minute, as the variations in the shear associated to the thermal disturbances are very small \citep{2017MNRAS.472.4204M}. Besides, as the length scale of the thermal disturbances is largely smaller than the pressure's length scale, modelling the effects of a weak turbulence by the use of a laminar viscosity might not be an appropriate procedure. However, we do include an artificial von Neumann-Richtmyer ``viscous pressure''  as described in \citet{2016ApJS..223...11B} as well as the resulting compressional heating, the effects of which are negligible over the short duration runs performed in the present study.

\section{Results from linear theory}
\label{sec:results-from-linear}
We summarise here only the results relative to an eccentric planet. Those relative to a planet on a circular orbit can be found in \citet{2017MNRAS.472.4204M} or \citet{2021MNRAS.501...24C}.

Depending on the value of the eccentricity, its time evolution can be described either using an expansion of the response of the disc to first order in the eccentricity \citep{2019MNRAS.485.5035F} or using a dynamical friction expression \citep{2017MNRAS.465.3175M}. The latter is valid when the response time of the drag force is shorter than the shear time scale. The response time, to within a factor of order unity, reads \citep{2017MNRAS.465.3175M}:
\begin{equation}
  \label{eq:3}
   \tau_\textrm{DF} = \frac{\chi}{V^2},
\end{equation}
where $V$ is the velocity of the planet with respect to the ambient gas, while the shear time scale is simply $\Omega^{-1}$. By requesting that $\tau_\mathrm{DF}\ll \Omega^{-1}$, and using $V\sim E\Omega$, where $E=ea$ is the epicyclic excursion, we are led to $E\gg E_c$, where:
\begin{equation}
  \label{eq:2}
  E_c = \sqrt{\frac{\chi}{\Omega}}
\end{equation}
is the critical value for the epicyclic excursion separating the two regimes: that of low eccentricity, or shear-dominated regime, and that of larger eccentricity, or headwind regime. Incidentally, this critical value is also the typical  size $\lambda_c$ of the thermal lobes that surround a planet on a (nearly) circular orbit \citep{2017MNRAS.472.4204M}, which has the expression:
\begin{equation}
  \label{eq:10}
  \lambda_c=\sqrt{\frac{\chi}{(3/2)\Omega\gamma}}.
\end{equation}
In the planet-forming regions of protoplanetary discs, this length scale is usually smaller than the pressure length scale $H$ by an order of magnitude.

In the shear-dominated regime,  \citet{2019MNRAS.485.5035F} find that the eccentricity growth time averaged over one orbit is:
\begin{equation}
\label{eq:e_dot}
    \bar{\dot{e}} = e\frac{1.46}{t_\text{thermal}}\left(\frac{L}{L_c}-1\right),
  \end{equation}
  where the critical luminosity $L_c$ is
\begin{equation}
\label{eq:Lc}
    L_c = \frac{4\pi G M_p\chi\rho_0}{\gamma},  
\end{equation} 
and where $t_\text{thermal}$ is the thermal time, defined as:
\begin{equation}
    t_\text{thermal}=\frac{c^2_s\Omega_p\lambda_c}{2(\gamma-1)G^2M\rho_0}.
\end{equation}
When compared to the damping time $t_\text{wave}$ of the adiabatic case\footnote{It may seem surprising to refer to the work of \citet{tanaka2002} or \citet{2004ApJ...602..388T} as representative of the adiabatic case, since they explicitly make the assumption of an isothermal gas. However, results about eccentricity damping come from the excitation of density waves and can be straight forwardly applied to the adiabatic case by substituting the isothermal sound speed $c_s^\mathrm{iso}$ with the adiabatic sound speed $c_s=\gamma^{1/2}c_s^\mathrm{iso}$.} \citep{2004ApJ...602..388T}, we find: 
\begin{equation}
\label{eq:t_ratio}
    \frac{t_\text{thermal}}{t_\text{wave}}=\sqrt{\frac{\pi}{2}}\frac{\lambda_c}{\gamma(\gamma-1)H}. 
\end{equation}
As $\lambda_c \ll H$, this implies that the eccentricity evolution of a planetary embryo is dominated by the thermal forces (unless $L\approx L_c$, as in such case the right hand side (R.H.S.) of Eq.~\eqref{eq:e_dot} is small.)

More specifically, we see on Eq.~\eqref{eq:e_dot} that if the embryo's luminosity $L$ is smaller than $L_c$, its eccentricity is damped on a timescale shorter than that of \citet{2004ApJ...602..388T}, whereas it grows exponentially if $L>L_c$.

In the headwind-dominated case, similar considerations apply. The force exerted on the embryo is directed along the motion when $L>L_c$, which leads to a growth of eccentricity \citep{2017arXiv170401931E}, and against the motion when $L<L_c$ \citep{2017MNRAS.465.3175M}, which leads to a damping of eccentricity. In this regime, however, the growth or decay of eccentricity is not exponential with time, and when $L>L_c$ the eccentricity eventually saturates to a value comparable to the disc's aspect ratio \citep{2017arXiv170401931E}.

Thermal effects have been found to be in agreement with those predicted by linear theory for planetary masses up to the critical mass:
\begin{equation}
M_c = \chi c_s/G,
\label{eq:criticalmass}
\end{equation}
and to decay past this critical mass, from numerical experiments in the headwind dominated regime \citep{2020MNRAS.495.2063V}.  For parameters typical of the planet forming regions of a protoplanetary disc at a few astronomical units, the decay is expected for planetary masses comparable to the Earth's mass. Our numerical exploration will therefore consider protoplanet masses from a fraction of an Earth mass to a few Earth masses.

\section{Numerical implementation}
\label{sec:numer-impl}
We make use of the version of \texttt{FARGO3D}\footnote{\texttt{https://bitbucket.org/fargo3d/public}} \citep{2016ApJS..223...11B,fargo2000} augmented with the nested meshes functionality described by \citet{2019MNRAS.483.4383V}. Several changes were implemented in order to enable the code to perform the high-resolution simulations needed to study the thermal interaction between planet and disc. In this section we go into some detail on the implementation of these changes, which cover the last two terms of Eq.~\eqref{eq:4}.  The corresponding contributions are treated in separation, using the operator splitting technique, while the rest of the equation is solved as detailed in \citet{2016ApJS..223...11B}.

\subsection{Implementation of thermal diffusion}
\label{sec:adding-therm-diff}
As done in \citet{2019MNRAS.483.4383V}, we make use of a simple modelling of the thermal diffusion process, considering a constant thermal diffusivity $\chi$. The numerical implementation starts by computing the energy fluxes multiplied by their corresponding surfaces as:
\begin{eqnarray}
  f^x_{i-\frac12} &=& \chi\left(\frac{\rho_i+\rho_{i-1}}{2}\right)\left(\frac{e_i}{\rho_i}-\frac{e_{i-1}}{\rho_{i-1}}\right)\frac{S_x}{\Delta x}\\
  f^y_{j-\frac12} &=& \chi\left(\frac{\rho_j+\rho_{j-1}}{2}\right)\left(\frac{e_j}{\rho_j}-\frac{e_{j-1}}{\rho_{j-1}}\right)\frac{S_y}{\Delta y}\\
  f^z_{k-\frac12} &=& \chi\left(\frac{\rho_k+\rho_{k-1}}{2}\right)\left(\frac{e_k}{\rho_k}-\frac{e_{k-1}}{\rho_{k-1}}\right)\frac{S_z}{\Delta z},
\end{eqnarray}
where half integer indices refer to interfaces and where integer indices with unique value $i$, $j$ or $k$ throughout the expression are omitted for legibility.
We then update the density of internal energy after a time step $\Delta t$ as:
\begin{equation}
  \label{eq:22}
	\mathfrak{e}_{i,j,k}^{n+a}= \mathfrak{e}_{i,j,k}^{n}-\frac{\Delta t}{V} \left(f^x_{i+\frac12}-f^x_{i-\frac12}+ f^y_{j+\frac12}-f^y_{j-\frac12}+f^z_{k+\frac12}-f^z_{k-\frac12} \right),
\end{equation}
where $V$ is the volume of the cell. Since the time step limit associated to this substep scales as $\min(\Delta x,\Delta y, \Delta z)^2/\chi$, the CFL constraint is eventually dominated by thermal diffusion for the highest levels of resolution. However, with our setup the time constraint arising from thermal diffusion never reduces significantly the timestep. In Eq.~\eqref{eq:22}, $a$ is an arbitrary variable meant to convey that the value obtained for the energy density is an intermediate one, which is later used in the substep dedicated to the heat release, as explained in the next section.

\subsection{Implementation of heat release}
\label{sec:adding-heat-release}
We use a simple implementation to emulate the heat release by the luminous planet into the disc. This implementation is similar to those used by \citet{2015Natur.520...63B}, \citet{2017arXiv170401931E} and \citet{2019MNRAS.483.4383V}. For the cells that comply with $|x_p-x_{i,j,k}|<\Delta_x$, $|y_p-y_{i,j,k}|<\Delta_y$ and $|z_p-z_{i,j,k}|<\Delta_z$, i.e. the cells nearest to the planet, the internal energy is increased as: 

\begin{equation}
  \label{eq:13}
    \mathfrak{e}_{i,j,k}^{n+1} = \mathfrak{e}_{i,j,k}^{n+a} + L\left(1-\frac{x^{'}_{i,j,k}}{\Delta_x}\right)\left(1-\frac{y^{'}_{i,j,k}}{\Delta_y}\right)\left(1-\frac{z^{'}_{i,j,k}}{\Delta_z}\right)\frac{\Delta t}{V},
  \end{equation}
  where $\xi^{'}_{i,j,k}=|\xi_p-\xi_{i,j,k}|$ for $\xi=x,y,z$.
  While an exact treatment of the heat released by the planet and of the temperature and radiation fields that it entails in its neighbourhood is considerably more complex than described by a diffusion equation with a point-like source term, we note that the mean Rosseland opacity in the planet forming regions of protoplanetary discs yields a photon mean free path that is comparable to the size of our smallest cells \citep[see e.g.][Tab.~1]{2017MNRAS.465.3175M}. The use of a diffusion equation should therefore correctly capture the properties of the heated region, while  there is no need for an implementation of heat release more sophisticated than that described by Eq.~\eqref{eq:13}, which corresponds to the  local release of the heat in the cells next to the planet.

We have validated our implementation of the heat release and the disc's thermal diffusion by running a low-mass planet on a circular orbit and checking that the perturbation induced in the gas by the heat released has the shape and amplitude predicted by \citet{2017MNRAS.472.4204M}. This validation procedure, presented in Appendix~\ref{app:1}, is similar to that of \citet{2020ApJ...902...50H} or that of \citet{2021MNRAS.501...24C}.
  
\subsection{Three-dimensional spherical nested meshes}
\label{sec:three-dimens-spher}
\begin{figure*}
    \centering
    \includegraphics[width=\textwidth]{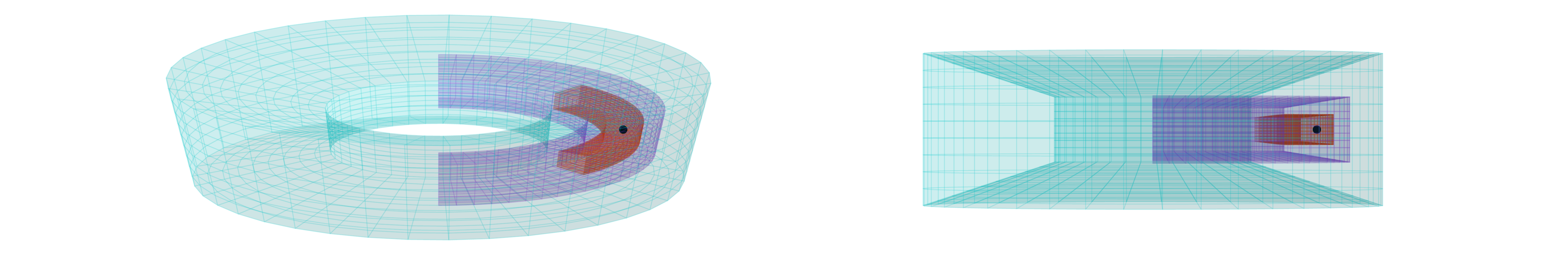}
    \caption{Visualisation of the 3D spherical nested meshes.}
    \label{fig:meshes}
\end{figure*}

We present a brief description of the $3D$ spherical nested meshes setup used in all simulations presented in this work. As described in \citet{2019MNRAS.483.4383V}, the \texttt{FARGO3D} code was redesigned to include a hierarchy of nested meshes or levels of refinement. In this hierarchy, each added level $\ell+1$ doubles the resolution in all three dimensions with respect to the preceding level $\ell$. In Fig.~\ref{fig:meshes} we present a visualisation of a system with two levels of refinement (for a total of three levels). The nested meshes are  centred on the planet, providing therefore the highest resolution of the simulation near the planet's location. This implementation of nested meshes provides  a fixed refinement, i.e. the position of the nested meshes is determined at the beginning of the simulation and remains unchanged throughout the simulation. Therefore, it is necessary to use a frame in corotation with the planet in order to keep the planet inside the refined region. 

\subsection{Non-fixed orbits in a nested-meshes system}
\label{sec:non-fixed-orbits}
In our simulations the planet is free to move according to the disc's force, rather than constrained to orbit on a fixed circular or eccentric path.
We therefore need to compute the force exerted by the disc onto the planet. This task is not as straightforward as it seems in the context of a disc described by a collection of nested meshes. The planet trajectory is updated with the smallest timestep, computing the force exerted onto the planet by the whole set of nested meshes. Each level takes its turn to be updated, starting from the highest (finest resolution) to the lowest (coarsest resolution). For each step taken by a mesh of level $\ell$, two steps are taken by a mesh of level $\ell+1$. This means that the system is fully synchronised upon completion of a time step $\Delta t$ on the ground level ($\ell=0$). We update the planet's velocity and position every $\Delta t/2^n$, where $n$ is the total number of nested meshes, i.e. for each timestep on the finest level. When the $i^{th}$ ($i\in [1,2^n]$) update on the finest level ($n$)  has completed, we also update level $n-1$ ($n-j$) if $i$ is a multiple of $2$ ($2^j$)
prior to the force evaluation, so that the levels are as synchronised as possible when evaluating the force.

\subsection{Fiducial Disc model}
\label{sec:fiducial-disc-model}
In Tab.~\ref{tab:1} we give the values of all the parameters of our fiducial disc model. Our code units are, for mass, that of the central star, for distance, the semi-major axis of the planet, and the time unit is chosen such that $G=1$,
which entails that the orbital period of the planet is $2\pi$. When converting our fiducial values to c.g.s. units, we specify to the case of central star of solar mass and a planet with a semi-major axis equal to that of Jupiter ($5.2$~au). The value of the thermal diffusivity $\chi$ that we have adopted is that evaluated by \citet{2017arXiv170401931E} in their fiducial disc. Using Eq.~\eqref{eq:10} we find:
\begin{equation}
    \label{eq:lambda_c}
    \lambda_c = 4.6\times 10^{-3} a = 0.09H,
\end{equation}
which is a small fraction of the pressure scale length \citep{2017MNRAS.472.4204M}.

\begin{table}
    \centering 
    \begin{tabular}{l|c|c}
        \hline 
        Parameter & Value in c.u.  &  Value in c.g.s. units \\
        \hline 
      $a$  & $1.0$ & $7.8\times 10^{13} \text{cm}$ ($5.2$ au) \\
      $\Omega_p$  & $1.0$ & $1.68\times 10^{-8}\text{s}^{-1} $ \\
      $\chi$ & $4.5\times10^{-5}a^2\Omega_p$  &  $4.6\times10^{15}\text{cm}^2/s$ \\
      $M_\star$ & $1.0$ & $2\times 10^{33}$g ($1M_\odot$)\\
      $\Sigma_0$ & $6.05 \times 10^{-4}M_\star/a^2$ & $200 g/\text{cm}^2 $\\
      $h(a)$ & $0.05$ & $0.05$\\
      $M_p$  &  $2.66\times 10^{-6}M_\star$  &  $5.32 \times 10^{26}g$ \\
        \hline 
    \end{tabular}
    \caption{\label{tab:1} Values for the fiducial disc used in all the simulations presented in section~\ref{sec:results}. We use Plummer's law for the planetary potential, with a softening length equal to $5\cdot 10^{-4}a$.}
    \end{table}{}
    
In Tab.~\ref{tab:mehses} we present the description of the levels of refinement, their number of cells and extent in the three dimensions of our coordinate system. We chose a frame corotating with the planet and not with its guiding centre so that the planet moves only in the radial direction, which allows us to narrow the extent of the nested meshes in the azimuthal direction. Therefore, in this setup, the limiting factor in terms of the number of nested meshes to use comes from the radial extent of the finest level, as we need this level to host the planet at all times. We have used a maximum of three levels of refinement, though most of the simulations where performed with two levels of refinement. This setup allows the planet to have for $\ell_\text{max} = 2$ an eccentricity with value up to $0.2$, and the resolution of the finest level to be $(\Delta \phi,\Delta r, \Delta \theta)=(0.106,0.108,0.108)\lambda_c$, whereas for $\ell_\text{max} = 3$ the eccentricity cannot exceed $0.1$, while the resolution of the finest level reaches $(\Delta \phi,\Delta r, \Delta \theta)=(0.053,0.054,0.054)\lambda_c$. The simulations with $\ell_\text{max} = 3$ were used solely for purposes of convergence study.

\begin{table}
    \centering 
    \begin{tabular}{l|c|c|c|c}
        \hline 
        $\ell$ & $(N_\phi,N_r,N_\theta)$ & $(\phi_\text{min},\phi_\text{max})$ & $(r_\text{min},r_\text{max})$ & $(\theta_\text{min},\theta_\text{max})$\\
        \hline 
        $0$ & $(800,400,60)$  &  $\pi/4(-1,1)$ &  $(0.6,1.4)r_p$ &  $(\frac{\pi}{2}-0.12,\frac{\pi}{2})$\\
        $1$ & $(400,600,30)$  &  $\pi/16(-1,1)$ &  $(0.7,1.3)r_p$ &  $(\frac{\pi}{2}-0.03,\frac{\pi}{2})$\\
        $2$ & $(400,800,30)$  &  $\pi/32(-1,1)$ &  $(0.8,1.2)r_p$ &  $(\frac{\pi}{2}-0.015,\frac{\pi}{2})$\\
        $3$ & $(400,800,30)$  &  $\pi/64(-1,1)$ &  $(0.9,1.1)r_p$ &  $(\frac{\pi}{2}-0.0075,\frac{\pi}{2})$\\
        \hline 
    \end{tabular}
    \caption{\label{tab:mehses}Description of the sizes and extension of each level of the system of nested meshes designed to host the simulations for the eccentric planet. Level 2 is designed so that it can host planets with $e<0.2$, level 3 is meant to be used only for $e<0.1$.}
  \end{table}{}
  The use of nested meshes allows to improve considerably the resolution of the simulations of \citet{2017arXiv170401931E}. In said work, the resolution achieved was $(\Delta \phi,\Delta r,\Delta \theta)\approx(1,1/3,1/4)\lambda_c$, hence barely sufficient to capture the hot, underdense region.

  We also note that the resolution for $\ell_\mathrm{max}=2$ corresponds to $10$ to $64$ cells over the width of the horseshoe region of the circular case, for the range of planet masses that we consider, or to $9$ to $32$ cells for the Hill radius.

\subsection{Boundary conditions}
Since we neglect the inclination, the planet can be considered to orbit in the disc midplane. We therefore simulate only the upper half of the disc and use symmetric boundary conditions at the equator for all fields, except the velocity in colatitude for which we use an antisymmetric boundary condition. At the radial boundaries and at the lower boundary in colatitude, we extrapolate the density, energy density and azimuthal velocity fields using the dependence of these fields on radius and colatitude for discs in hydrostatic equilibrium, while we take symmetric boundary conditions for the colatitude (radial) component of the velocity in the radial (colatitude) buffer zones, and antisymmetric boundary conditions for the radial (colatitude) component of the velocity in the radial (colatitude) buffer zones. In addition, we apply wave-killing boundary conditions \citep{valborro06} in radius (from $r_\mathrm{min}$ to $r_\mathrm{min}\times 1.15^{2/3}$ and from $r_\mathrm{max}/1.15^{2/3}$ to $r_\mathrm{max}$) and at the lower limit in colatitude, (from $\pi/2-0.12$ to $\pi/2-0.096$), with a quadratic ramp and a damping time equal to $0.3/(2\pi)$ times the local Keplerian period. This damping procedure is performed for all variables (density, velocities and energy density).

\section{Results}
\label{sec:results}
\begin{figure*}
    \centering 
    \includegraphics[width=0.8\textwidth]{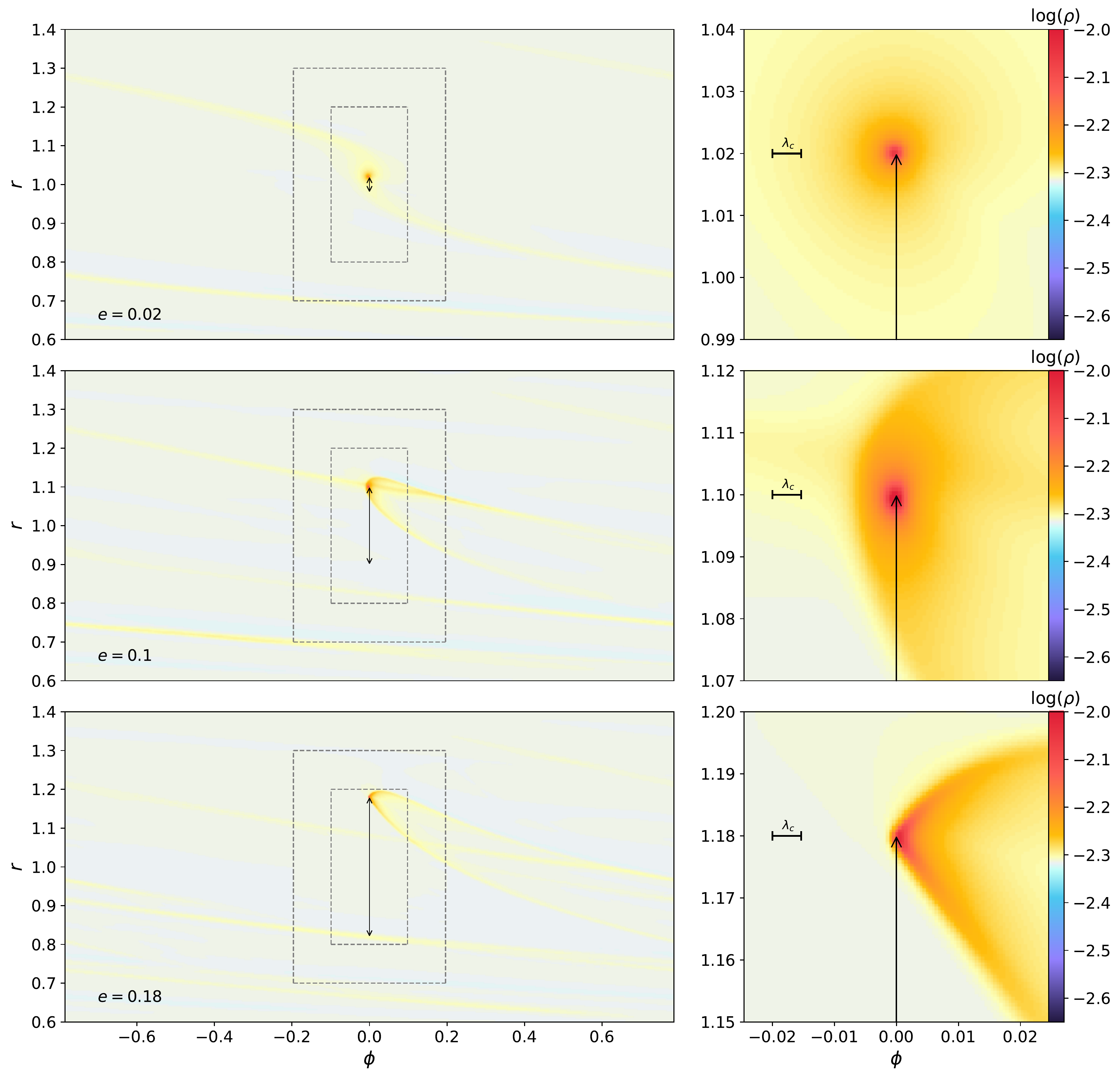}
    \caption{Snapshots of the density perturbation after 2 orbits for a system with 3 levels of refinement (as described by Tab.~\ref{tab:mehses}). On the left the complete system of nested meshes, where the grey dashed lines delimit the nested meshes, whereas the black line with arrows represents the planet's path in the corotation frame, which amounts to a purely radial oscillation. On the right the maximum level of refinement. The planet has a mass $M=M_\oplus$, a luminosity $L=2L_c$ for three different values of the eccentricity of its orbit $e=0.02,0.1$ and $0.18$.}
    \label{fig:slice} 
\end{figure*}

\begin{figure*}
    \centering 
    \includegraphics[width=0.8\textwidth]{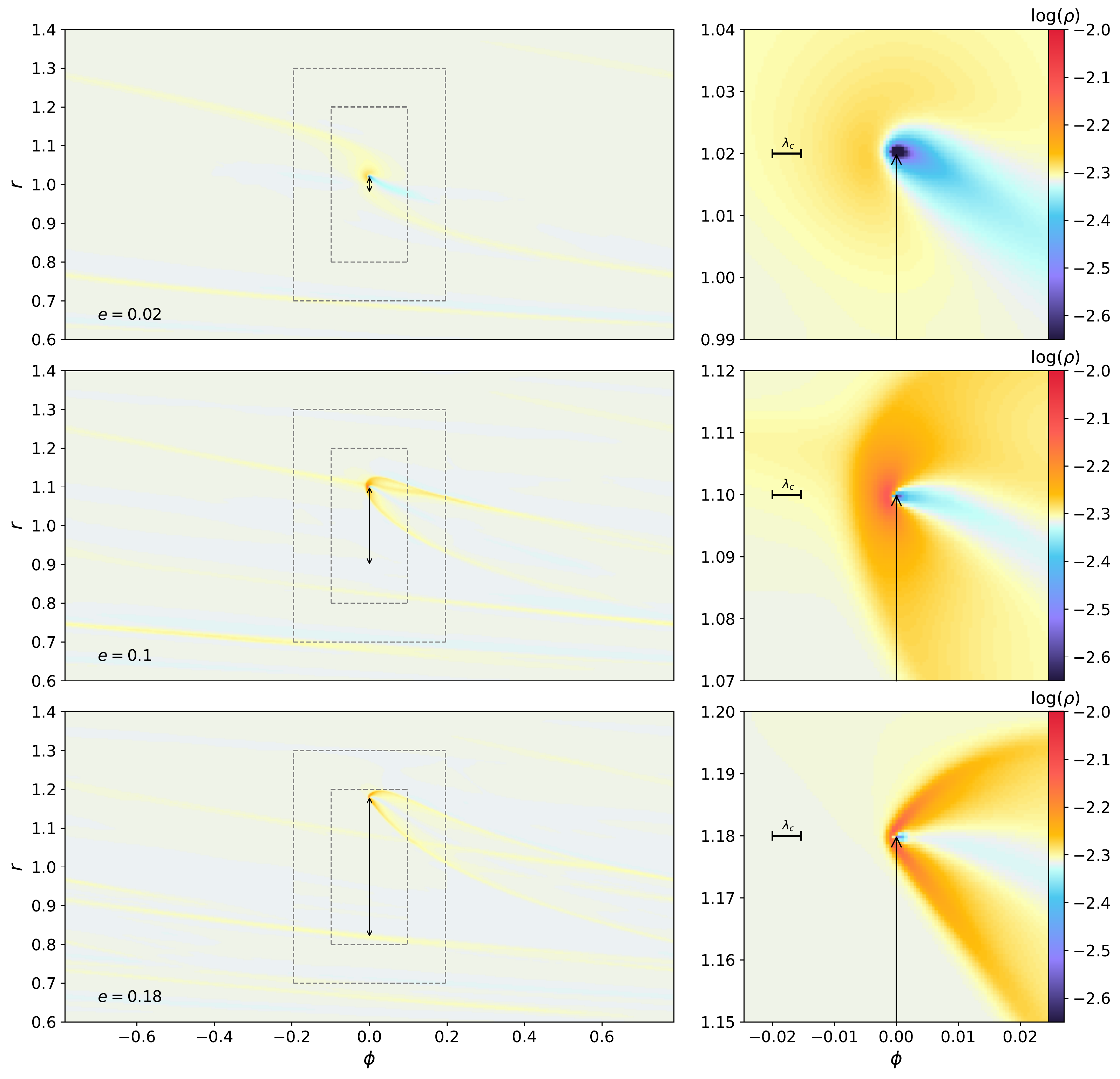}
    \caption{Snapshots of the density perturbation after 2 orbits for a system with 3 levels of refinement (as described by table~\ref{tab:mehses}). On the left the complete system of nested meshes, where the grey dashed lines delimit the nested meshes, and in a black line we present the planet's path in the corotation frame. On the right the maximum level of refinement. The planet has a mass $M=M_\oplus$, a luminosity $L=8L_c$ for three different values of the eccentricity of its orbit $e=0.02,0.1$ and $0.18$.}
    \label{fig:slice2} 
\end{figure*}

All our runs consist of short term simulations (over 3 orbital periods), for planets of a given mass $M_p$, of a given, constant luminosity $L$ and of a non-zero initial eccentricity $e$. From these runs we evaluate the time averaged eccentricity growth rate. We therefore obtain samples of the function:
\begin{equation*}
  (M_p,L,e)  \mapsto  \dot e(M_p,L,e),
\end{equation*}
for our given disc model and at the locations in parameter space $(M_p,L,e)$ corresponding to our simulations.
Since we are interested in planets that become eccentric under the action of thermal forces, we limit ourselves to sufficiently large luminosities, in excess of $L_c$. Under these circumstances, as found by \citet{2017arXiv170401931E}, we find that when the initial eccentricity of a planet is sufficiently small, it tends to grow, whereas if the initial eccentricity is large enough it tends to decay. In between, there is a (unique) value of the initial eccentricity where we observe that the eccentricity remains constant. We call this eccentricity the asymptotic eccentricity, and denote it with $e_\infty$,  as it is the eccentricity that the planet reaches at large time regardless of its initial eccentricity\footnote{We make here the reasonable assumption that the time derivative of the eccentricity depends only on its present value, and not on its previous evolution. The findings of \citet{2017arXiv170401931E} are compatible with this assumption.}. We show on Figs.~\ref{fig:slice} and~\ref{fig:slice2} the disc response to an embedded Earth mass planet with various eccentricities, respectively for the case $L=2L_c$ and $L=8L_c$. We see how the extent of the most resolved level covers the radial path of the planet.

\subsection{Convergence study}
\label{sec:convergence-study}
\begin{figure}
    \centering
    \includegraphics[width=\columnwidth]{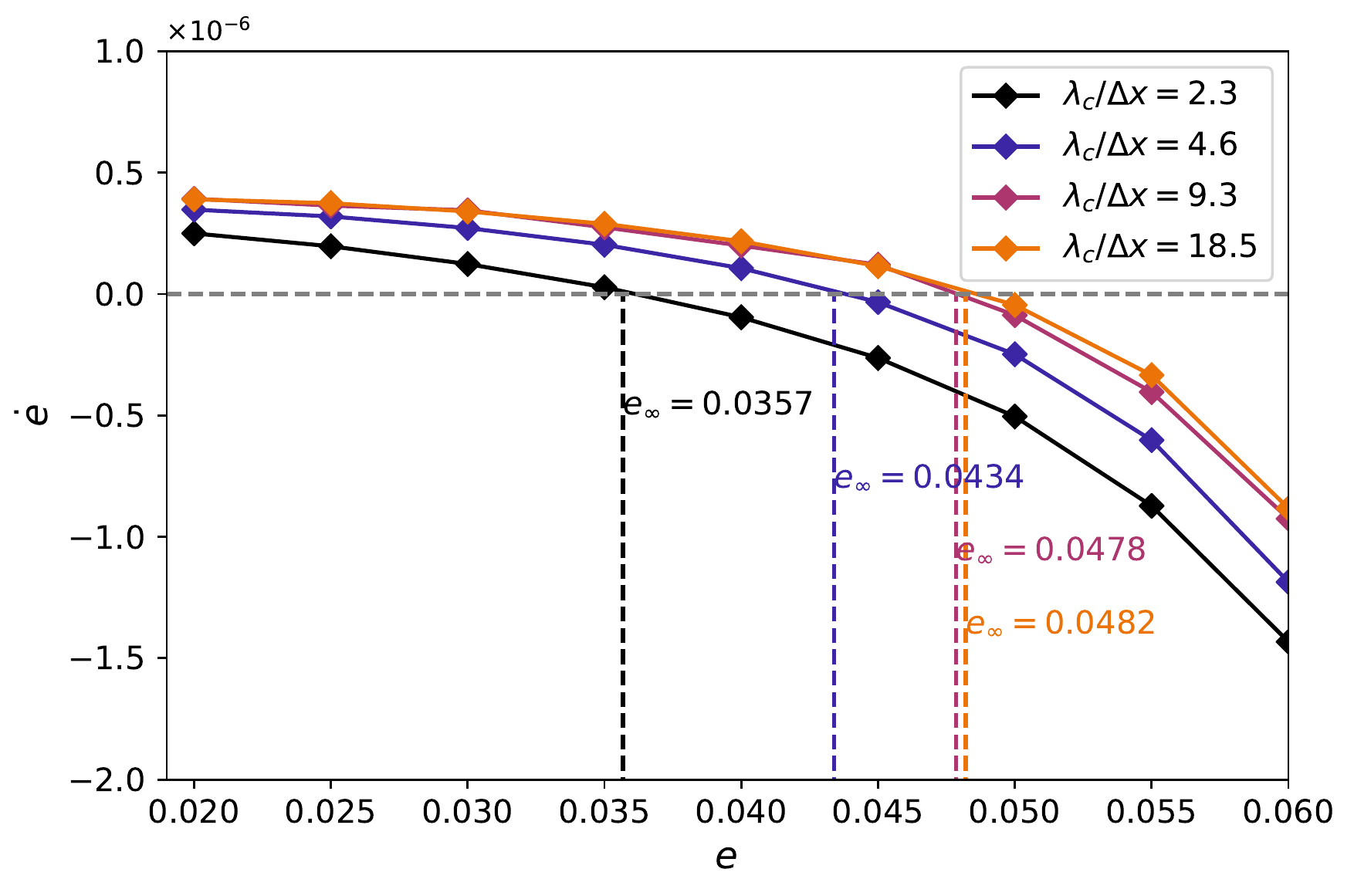}
    \caption{Eccentricity growth rate as a function of the eccentricity for different resolutions. The vertical dashed lines show the asymptotic eccentricities for each resolution.}
    \label{fig:convergence}
  \end{figure}

We start by presenting a study of the convergence of the asymptotic
eccentricity as a function of the resolution, specifically, as a function of the number of cells inside the characteristic length $\lambda_c$. The study was performed for a planet mass $M_p=0.1M_\oplus$ and luminosity $L=2L_c$.
This choice provides the shortest simulation time. Indeed the temperature, hence the sound speed which limits the timestep, increase with the luminosity in the vicinity of the planet. In our parameter space where the luminosity sweeps an interval given in terms of $L_c$, the luminosity increases with the mass,  since $L_c\propto M_p$, as shown by Eq.~\eqref{eq:Lc}.  Fig.~\ref{fig:convergence} shows the rate of eccentricity growth $\dot{e}$ for eccentricities in the range between $0.02$ and $0.06$, and for $\lambda_c/\Delta x = 2.3,4.6,9.3,18.5$. 

From these five curves we can extract the asymptotic eccentricity as the value of the eccentricity at which $\dot{e}=0$. In table~\ref{tab:2}, we present the asymptotic eccentricity as a function of resolution. We find that a resolution of $\Delta x=9.3/\lambda_c$ suffices to obtain a reliable estimate of the asymptotic eccentricity $e_\infty$ (it turns out to be $10.3\%$ larger than for half that resolution, and only $0.8\%$ smaller than for twice that resolution).
\begin{table}
    \centering 
    \begin{tabular}{l|c|c|c}
        \hline 
         $\lambda_c/\Delta x$ & $e_\infty$ & Increment in $\lambda_c/\Delta x$ & Increment in $e_\infty$\\
        \hline 
        $2.3$  & $0.0357$ & -         & - \\
        $4.6$  & $0.0434$  & $\times2$ & $+21.6\%$ \\
        $9.3$  & $0.0478$ & $\times2$ & $+10.3\%$ \\
        $18.5$ & $0.0482$ & $\times2$ & $+0.8\%$ \\
        \hline 
    \end{tabular}
    \caption{\label{tab:2}Results for the convergence study of the asymptotic eccentricity as a function of the resolution.}
\end{table}{}
We mention that we have also performed additional calculations with one nested mesh (instead of two nested meshes) for our fiducial resolution $\lambda_c/\Delta x = 9.3$, and found that the resulting eccentricity was $1.6\%$ smaller than in the calculation with nested meshes. Therefore, the quoted asymptotic eccentricities here on present a slight overestimation.
We note that the requirement that the size of the heated region being of the order of ten cells is in agreement with the study of \citep{2021MNRAS.501...24C}, who obtain a similar value for thermal forces exerted on planets on circular orbits.

\subsection{Asymptotic eccentricity for different planet masses}
\label{sec:asympt-eccentr-diff}
\begin{figure*}
    \centering
    \includegraphics[width=.95\textwidth]{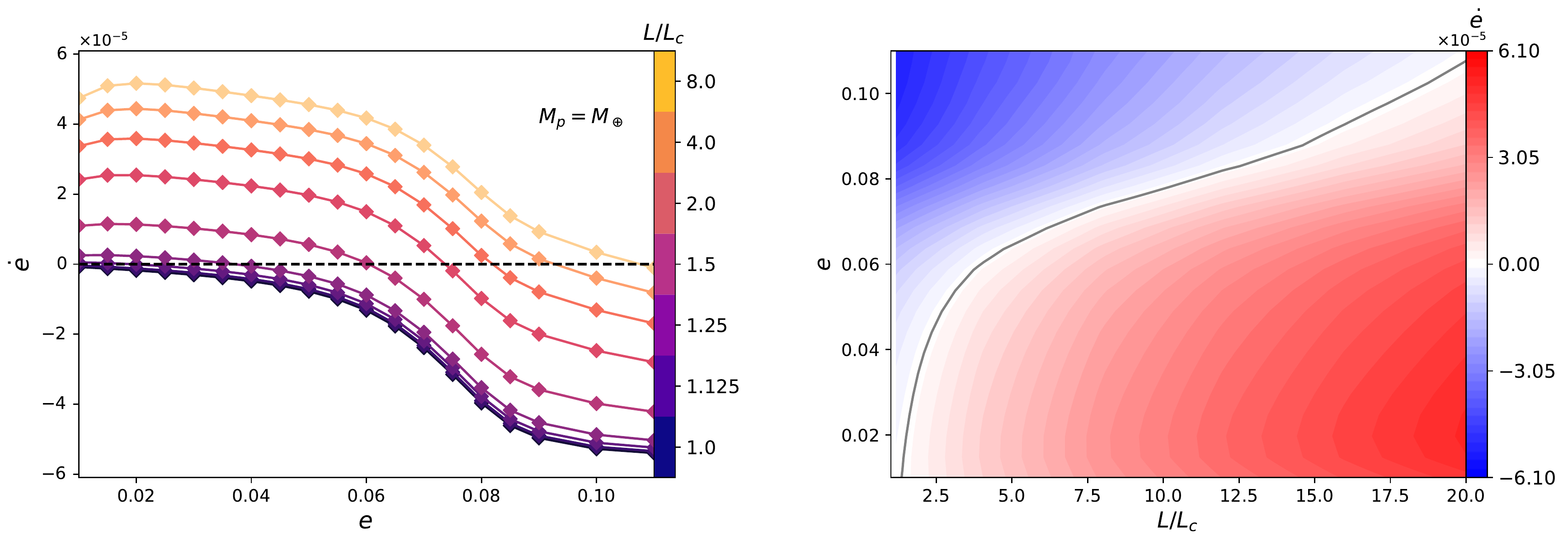}
    \caption{Left: eccentricity growth rate $\dot{e}$ as a function of the eccentricity $e$ for an Earth-mass planet for different luminosities. Right: interpolation of the results for $\dot{e}$ over the range of eccentricities and luminosities used. In red we present the region for $\dot{e}>0$ whereas in blue the region for $\dot{e}<0$. The solid line shows $\dot{e}=0$.}
    \label{fig:e_growth}
\end{figure*}

\begin{figure}
    \centering
    \includegraphics[width=\columnwidth]{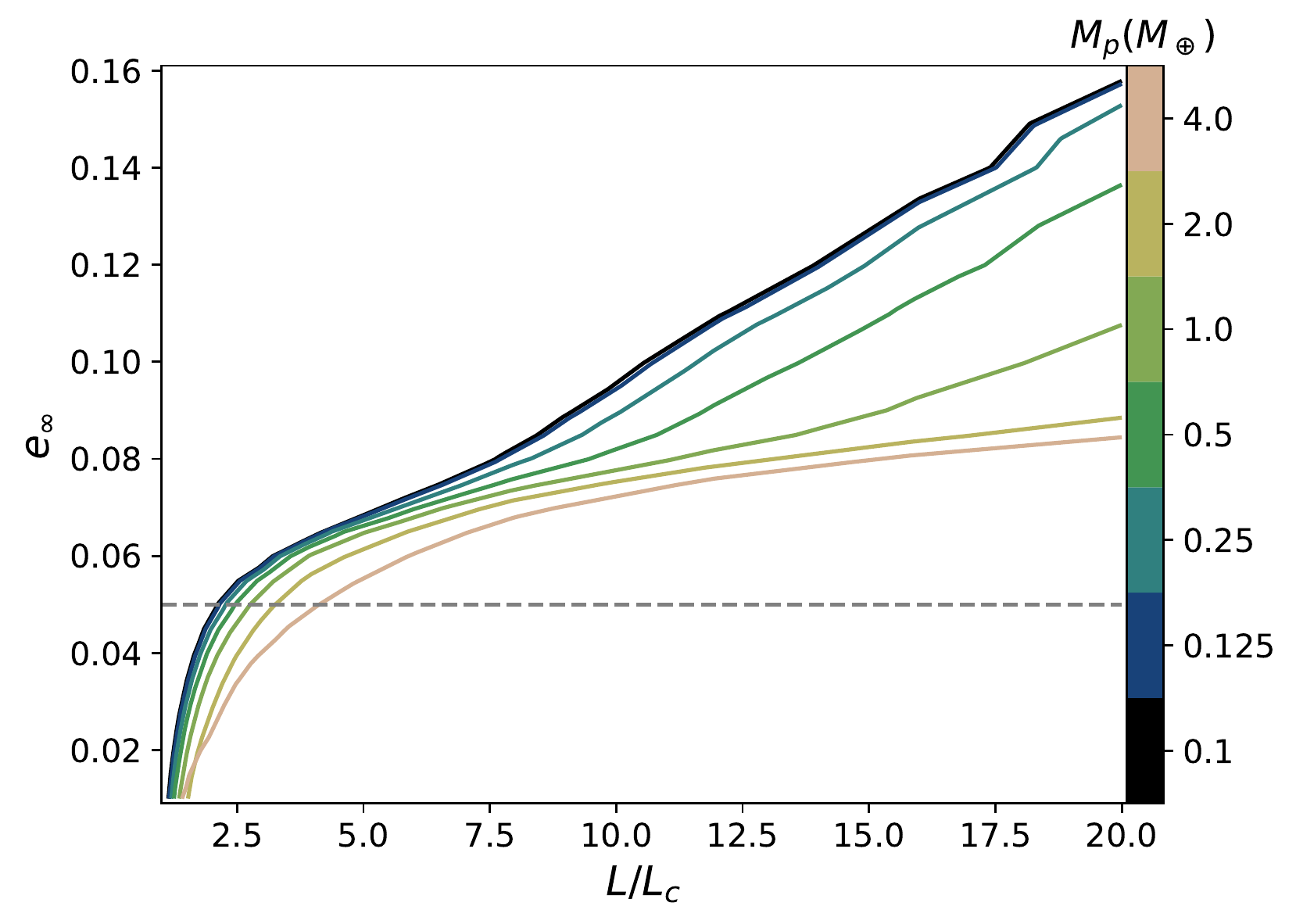}
    \caption{Asymptotic eccentricity as a function of the planet's luminosity, for different planet masses. The asymptotic eccentricity $e_\infty$ is obtained interpolating the results of $\dot e$ vs $e$ to the eccentricity for which the growth rate $\dot{e}$ would be zero. In a grey dashed line we present $h=0.05$ (the value used in the fiducial model for the aspect ratio). The third curve from the bottom (1~$M_\oplus$) corresponds to the black curve of Fig.~\ref{fig:convergence}.} 
    \label{fig:e_L}
\end{figure}

The results here and for the rest of the paper were obtained with a resolution of $\Delta x=9.3/\lambda_c$ in a system with two levels of refinement (as described in Tab.~\ref{tab:mehses}). The problem at hand has a $3D$ parameter space $(e, M_p/M_\oplus, L/L_c)$. We restrict our numerical exploration to $e \in [0.02,0.19]$, $M_p \in [0.1,4]M_\oplus$ and $L \in [1,36]L_c$. In Fig.~\ref{fig:e_growth} we present results for $M_p=M_\oplus$. The left plot shows $\dot{e}$ as function of $e$ for different values of $L/L_c$. On the right we present a colour map of $\dot{e}(e,L/L_c)$, where the solid line corresponds to  $\dot{e}(e,L/L_c)=0$.
From the discussion at the beginning of section~\ref{sec:results}, this curve also corresponds to the relationship between the planet's luminosity and its asymptotic eccentricity.
A similar numerical exploration was performed for $M_p=0.1,0.125,0.25,0.5,2$ and $4\;M_\oplus$. We omit presenting the analogues of Fig.~\ref{fig:e_growth} for these masses, as the information can be summarised in the graphs of $e_\infty$ vs. $(L/L_c)$ presented in Fig.~\ref{fig:e_L}. We observe that planets of larger mass reach a smaller asymptotic eccentricity for a given value of $L/L_c$. Lower planetary masses ($M_p<M_\oplus$) reach an
asymptotic eccentricity larger than the disc's aspect ratio for luminosities in excess of $3L_c$.

\subsection{Dependence on the aspect ratio}
\label{sec:depend-aspect-ratio}
\begin{figure*}
    \centering 
    \includegraphics[width=0.9\textwidth]{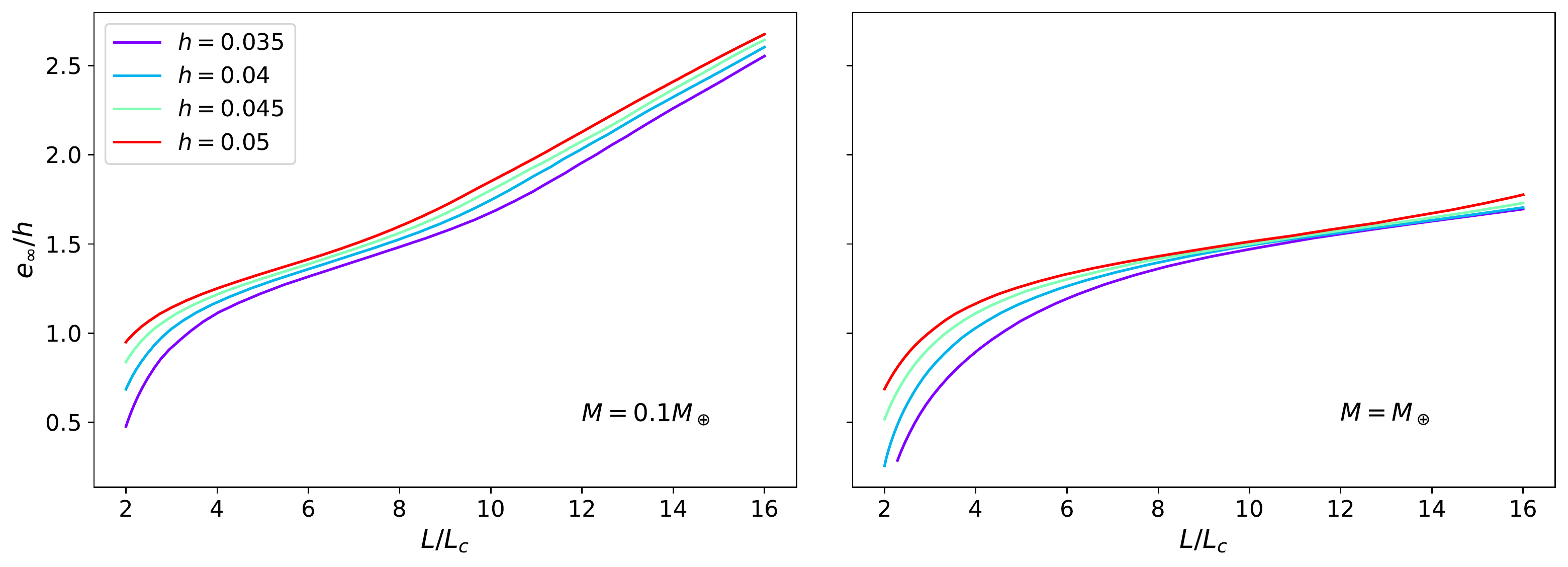}
    \caption{Ratio of the asymptotic eccentricity $e$ to the aspect ratio $h$ as a function of the planet's luminosity for various values of the disc's aspect ratio. The left plot shows results for the sub-critical mass $M=0.1M_\oplus$ and the right plot shows the results for a critical mass $M=M_\oplus$. 
      Except for the left part of the plots, we see that the different curves are nearly superimposed, which implies that the eccentricity scales with the aspect ratio.}
    \label{fig:aspectratio}
\end{figure*}

The results presented so far were all obtained for a disc's aspect ratio $h=0.05$.  The asymptotic eccentricity is attained when the growth rate, given by the thermal forces, is exactly compensated by the damping rate arising from the resonant interaction with the disc. The latter is given by a sum on the first-order Lindblad resonances \citep{arty93b,2004ApJ...602..388T} when the epicyclic excursion is much smaller than the pressure length scale, while it requires to incorporate higher order Lindblad resonances when this condition is no longer satisfied \citep{paplar2000}.
We expect the growth of the eccentricity to be limited past the aspect ratio, since thermal forces decay sharply when the perturber becomes supersonic \citep{2017MNRAS.465.3175M,2019MNRAS.483.4383V,2020MNRAS.495.2063V}, whereas the damping by resonant interaction continues to grow with eccentricity \citep{2007A&A...473..329C}. We therefore expect to observe smaller asymptotic eccentricity $e_\infty$ for smaller aspect ratios of the disc. We performed a study of the asymptotic eccentricity for two planetary masses ($M_p = 0.1$ and $1M_\oplus$), for three additional values of the aspect ratio ($h=0.035,0.04$ and $0.045$), the embryos' luminosities spanning  $L \in [2,16]L_c$. In Fig.~\ref{fig:aspectratio} we presents the results of this study. They show that when the
luminosity is much larger than $L_c$, the eccentricity is nearly proportional to the aspect ratio, for the two planet masses considered here.

\section{feedback of the eccentricity on the luminosity}
\label{sec:feed-back-eccentr}
Up to this point we have obtained, for our fiducial disc model, the eccentricity at larger time of a low-mass planet as a function of its mass and luminosity. The planet's luminosity, in turn, is mostly provided by the rate at which it accretes solids, either pebbles or planetesimals. In the following we will neglect that due to impacting planetesimals and we will focus exclusively on the luminosity arising from the accretion of pebbles.
The efficiency of pebble accretion represents the ratio between the available inward mass flux of pebbles $\dot{M}_\textrm{pebbles}$ and the rate at which the planet is accreting them $\dot{M}_p$:
\begin{equation}
    \epsilon_{2D} = \frac{\dot{M}_p}{\dot{M}_\textrm{pebbles}}. 
  \end{equation}
The efficiency of pebble accretion has been found to depend on the planet's eccentricity \citep{2018A&A...615A.138L}.  Broadly speaking, the efficiency of pebble accretion increases with the eccentricity when the latter is smaller than the disc's aspect ratio, then decays as the eccentricity keeps increasing (and may increase again at large eccentricity). There is therefore a peak of efficiency for values of the eccentricity comparable to the disc's aspect ratio. We see that there exists a feedback loop: as we saw in section~\ref{sec:asympt-eccentr-diff}, the eccentricity depends on the planet luminosity (and increases with the latter). In turn, the luminosity depends on the eccentricity: as the planet's eccentricity increases, it is able to accrete more pebbles and become more luminous. Our aim in this section is therefore to find the asymptotic eccentricity and luminosity of a given planetary embryo as a function of the underlying disc parameters.

\subsection{Asymptotic luminosity of a pebble-accreting planet}
\label{sec:asympt-lumin-pebble}
We consider the planet's luminosity to be:
\begin{equation}
  \label{eq:15}
    L = \frac{GM_p\dot{M}_p}{R_p}, 
  \end{equation}
  where $R_p$ is the planet's physical radius. This oversimple approximation will be discussed in section~\ref{sec:disc-concl}.

  Following \citet{2018A&A...615A.138L} and \citet{2018A&A...615A.178O} we calculate the efficiency of pebble accretion as a function of the eccentricity of the planet's orbit $e$, the headwind velocity of the gas (with respect to the circular, Keplerian motion at same location), the dimensionless stopping time $\tau_s$ of the pebbles and the vertical turbulence diffusivity $\alpha_z$.  We therefore obtain the accretion as a function of these parameters and of the inward mass flux of pebbles $\dot{M}_\textrm{accretion}(e,v_\textrm{headwind},\tau_s,\dot{M}_\textrm{pebbles},\alpha_z)$.

\begin{figure}
    \centering
    \includegraphics[width=\columnwidth]{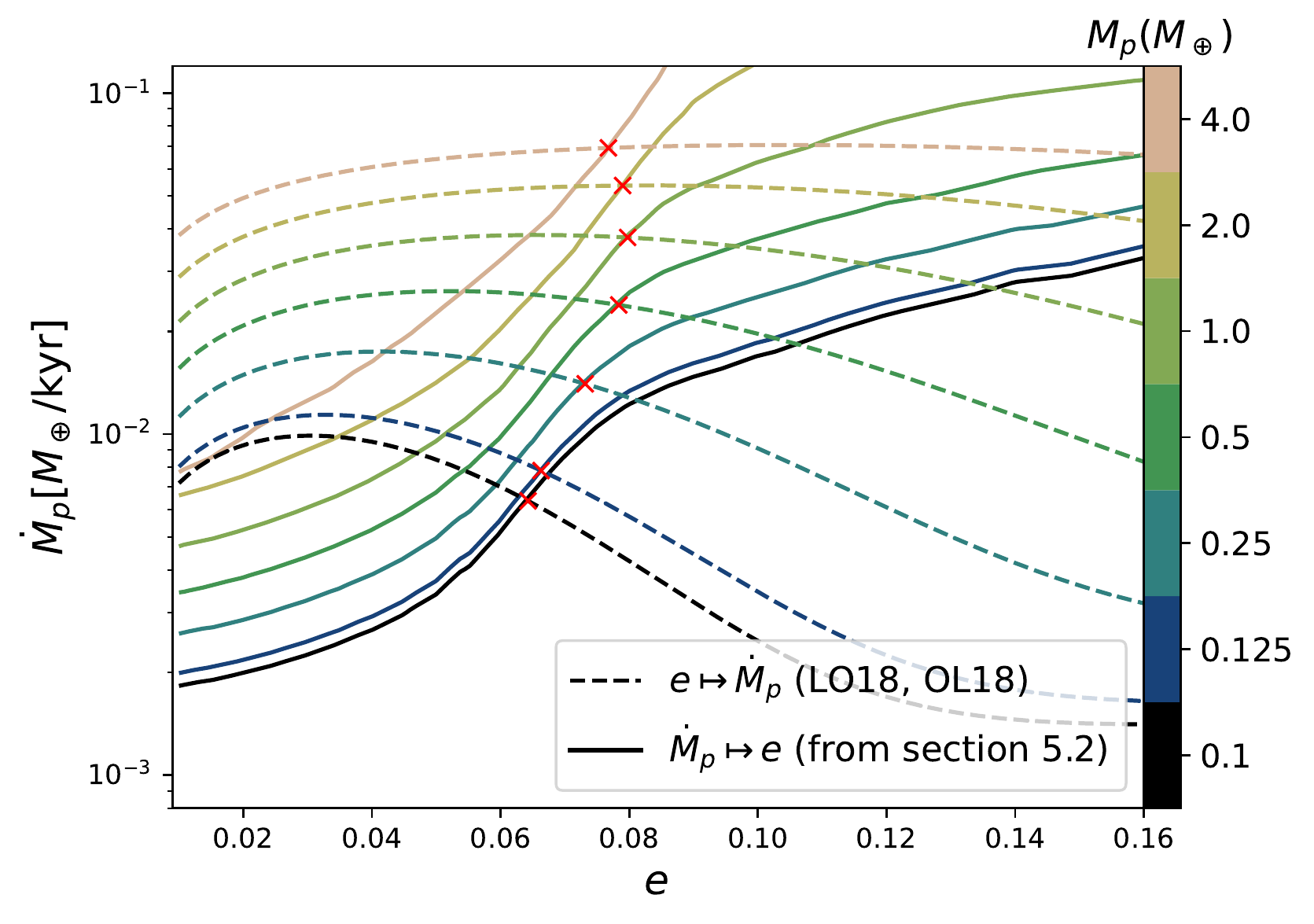}
    \caption{Intersection between the curves that give the accretion rate as a function of the eccentricity (from \citet{2018A&A...615A.138L,2018A&A...615A.178O}, dashed lines) and the curves that give the eccentricity reached for a given accretion rate, obtained from the parameter space exploration of section~\ref{sec:asympt-eccentr-diff} and Eq.~\eqref{eq:15} (solid lines).  The disc model has $v_\textrm{headwind}=26.2\textrm{m}/\textrm{s}$, $\tau_s=0.01$ and $\dot{M}_\textrm{pebbles}=10^{-4}M_\oplus\textrm{yr}^{-1}$. The red x marks show the intersection points for the different planetary masses.}
    \label{fig:intersection}
\end{figure}

For planetary cores of given density, the functions $L\mapsto e$ shown in Fig~\ref{fig:e_L} can be cast as functions $\dot M_p\mapsto e$, using Eq.~\eqref{eq:15}. Reciprocally, the work of \citet{2018A&A...615A.138L} and \citet{2018A&A...615A.178O} provide the accretion rate as a function of the eccentricity, $e\mapsto \dot M_p$, for a given set of disc parameters. Therefore, the value for the eccentricity reached at larger time by an embryo of a given mass in such disc may be found at the intersection point of the graphs of these two dependencies, as shown in Fig.~\ref{fig:intersection}.
That figure shows the results obtained for a given set of disc and pebbles parameters (the disc parameters others than those quoted in the caption being those of Tab.~\ref{tab:1}).
  We see that the largest eccentricity is $\sim 0.08$, reached by a critical mass planetary embryo (i.e. with an Earth mass), whereas that reached by the lowest mass planets considered here (with one tenth of an Earth mass) is $\sim 0.06$, still in excess of the aspect ratio, which indicates that in such case the planet still has a supersonic motion with respect to the underlying gas over a sizeable part of its epicycle.

\subsection{Dependence on the pebble's size and headwind}
\label{sec:depend-pebbl-size}
\begin{figure*}
    \centering
    \includegraphics[width=0.9\textwidth]{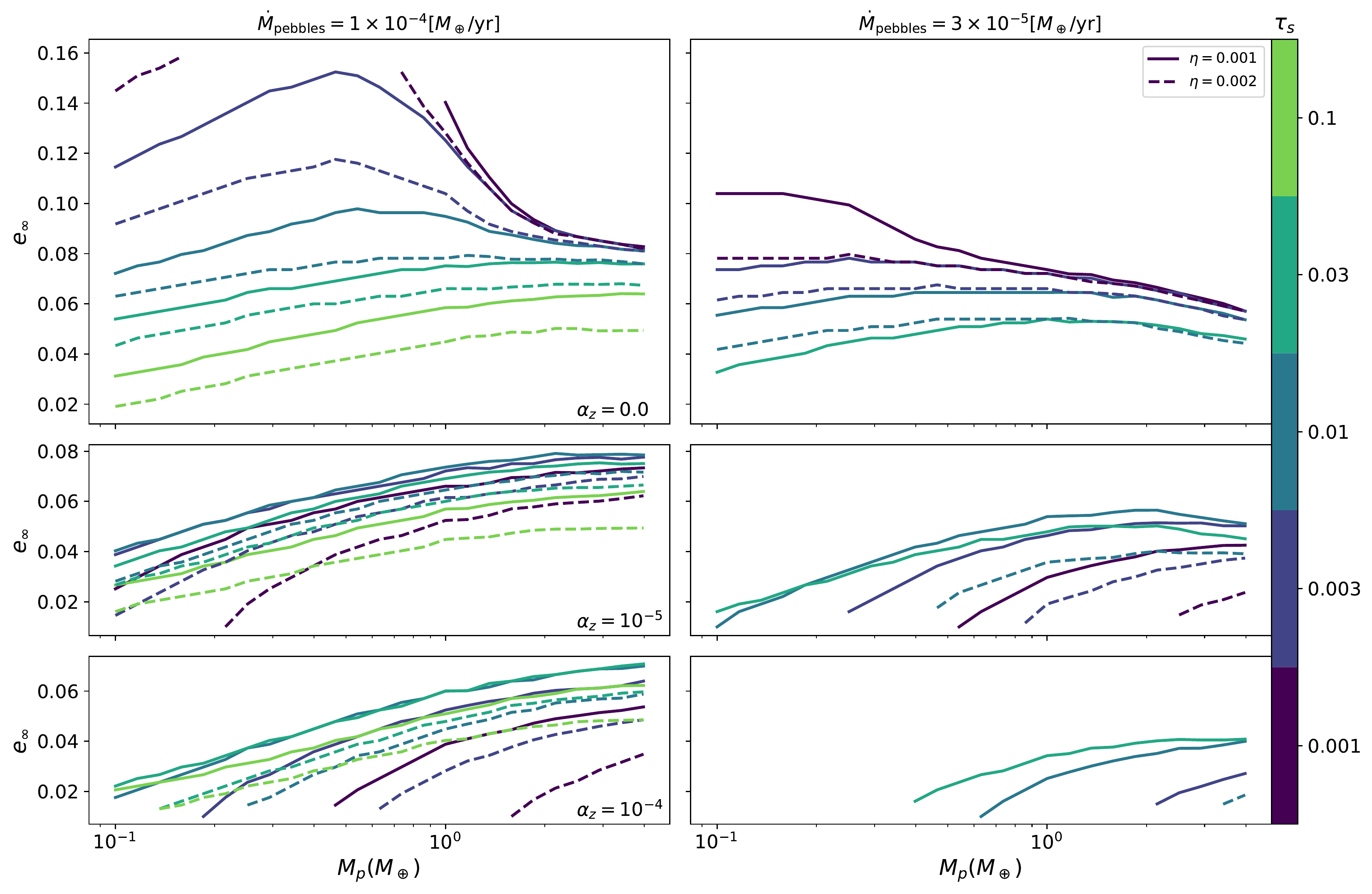}
    \caption{Asymptotic eccentricity as a function of the planetary mass for five different values of the dimensionless stopping time $\tau_s$ and for two different values of the headwind, for a mass flux of pebbles of $10^{-4}\;M_\oplus.\textrm{yr}^{-1}$ (left) and  $3\cdot 10^{-5}\;M_\oplus.\textrm{yr}^{-1}$ (right). These results show the high susceptibility of the asymptotic eccentricity to the size of the pebbles, their mass flux and the headwind velocity.}
    \label{fig:e_M}
\end{figure*}

As seen in section~\ref{sec:asympt-lumin-pebble}, the accretion rate is not only a function of the planet's eccentricity, it is also impacted by the available flux of pebbles, the size of the pebbles, the mismatch of orbital velocity between the planet and the gas, as well as the gas diffusivity. In Fig.~\ref{fig:e_M} we present the asymptotic eccentricity as a function of the planet's mass, for five different values of the dimensionless stopping time $\tau_s$ in the interval $[0.003,0.03]$, for two different values of the headwind velocity $v_\textrm{headwind}= \eta v_K$ with $\eta=0.001$ or $\eta=0.002$, which leads $v_\mathrm{headwind}=13.1\mbox{~or~}26.2 \textrm{m}/\textrm{s}$, for two different values of the mass flux of pebbles $\dot{M}_\textrm{pebbles}=5,10\times10^{-5}\;M_\oplus\mathrm{yr}^{-1}$, and for three values of the gas diffusivity $\alpha_z=0,10^{-5},10^{-4}$. Overall we see that the eccentricity has broadly same order of magnitude as the aspect ratio. We observe that as we decrease the value of the stopping time, the asymptotic eccentricity increases: smaller pebbles lead to larger eccentricities, as small pebbles drift more slowly and have a higher probability of being accreted by the planet. This results in higher accretion rates for a given mass flux. We also observe that for the larger values of the stopping time $\tau_s$ the asymptotic eccentricity increases with the planetary mass for sub-Earth masses, and reaches a roughly constant value beyond an Earth mass (for the largest of the two pebble fluxes) or slightly decays past an Earth mass (for the smallest pebble flux). In contrast, for smaller values of the stopping time $\tau_s$ there is a peak in eccentricity at around $0.5M_\oplus$, while the eccentricity decreases for larger masses. Therefore, the size of the pebbles has a significant impact on the asymptotic eccentricity for low-mass planets, whereas the impact is considerably smaller for high-mass planets. We observe that the mass at which the eccentricity peaks is larger for the largest pebble flux. We also observe that although the impact of the headwind velocity is similar in magnitude to that of $\dot{M}_\textrm{pebbles}$, it does not change the mass at which the eccentricity peaks. A smaller headwind velocity is found to favour larger eccentricities. This suggests that in dust traps, where the headwind velocity can take arbitrarily small values, and accretion rates can be much larger than in the rest of the disc, sub Earth-mass planetary embryos may reach large eccentricities. Finally, we observe that the gas diffusivity can also hinder eccentricity growth. The effect is observed to be large for small stopping times (highly coupled particles), and for smaller planetary masses. The impact seems to be negligible for larger stopping times (weakly coupled particles), as observed for the largest pebble inflow shown. We observe that for a turbulent flow ($\alpha_z>0$), the largest attainable eccentricities are found at larger masses, in contrast with the case of laminar flow, for which the planetary mass at which the eccentricity peaks varies with the pebble inflow rate and with the stopping time.

\begin{figure*}
    \centering
    \includegraphics[width=0.9\textwidth]{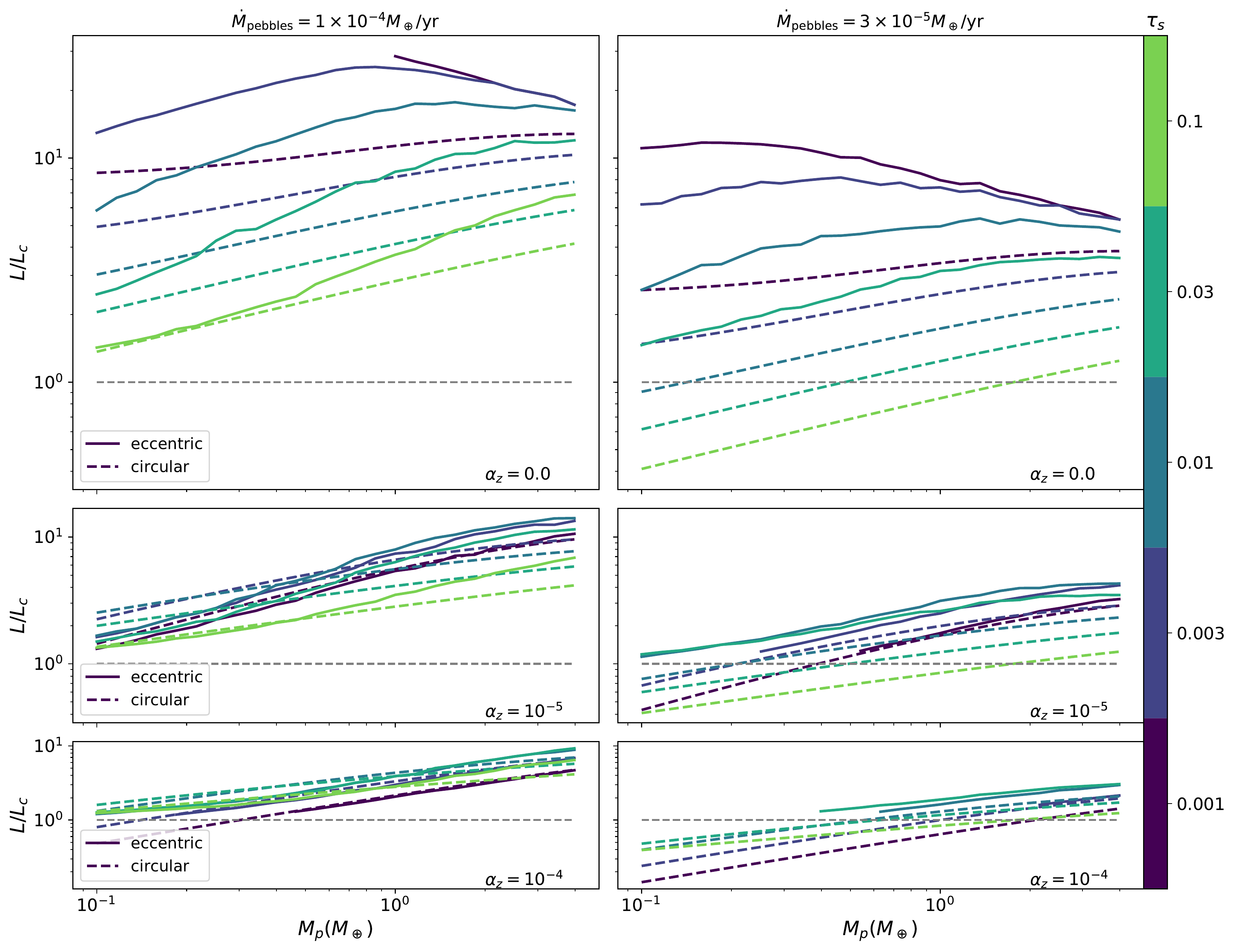}
    \caption{Luminosity as a function of the planetary mass for $\eta=0.002$, for five different values of the dimensionless stopping time $\tau_s$, for a mass flux of pebbles of $10^{-4}\;M_\oplus.\textrm{yr}^{-1}$ (left) and $3\cdot 10^{-5}\;M_\oplus.\textrm{yr}^{-1}$ (right). In solid lines we show the results obtained from the model developed in section~\ref{sec:feed-back-eccentr}, and in dashed lines we show the results obtained if the planet remains circular. These results show the potentially large impact that eccentricity has on the planet luminosity.}
    \label{fig:L_M}
\end{figure*}

Our models cannot only provide the asymptotic eccentricity as a function of the mass, but also the attainable pebble accretion rate as a function of the planet mass.  In Fig.~\ref{fig:L_M} we present the luminosity achieved as a function of the planet's mass, for the disc models presented in Fig.~\ref{fig:e_M}, focusing on the largest of the two values chosen for the head wind ($v_\textrm{headwind}= 26.2\; \textrm{m}/\textrm{s}$).  We also compare our results to that obtained assuming the planet remains on a circular orbit. We observe that for low $\alpha_z$, the planet reaches a luminosity (or accretion rate) systematically larger than the one it would have if it remained on a circular orbit.
As the turbulence strength parameter increases, the luminosity (or accretion rate) decreases, especially for subcritical mass planets, and may be comparable to the luminosity (or accretion rate) achieved on a circular orbit for $\alpha_z=10^{-4}$. For supercritical masses we recover the behaviour observed for a low turbulence strength.

\begin{figure*}
    \centering
    \includegraphics[width=0.9\textwidth]{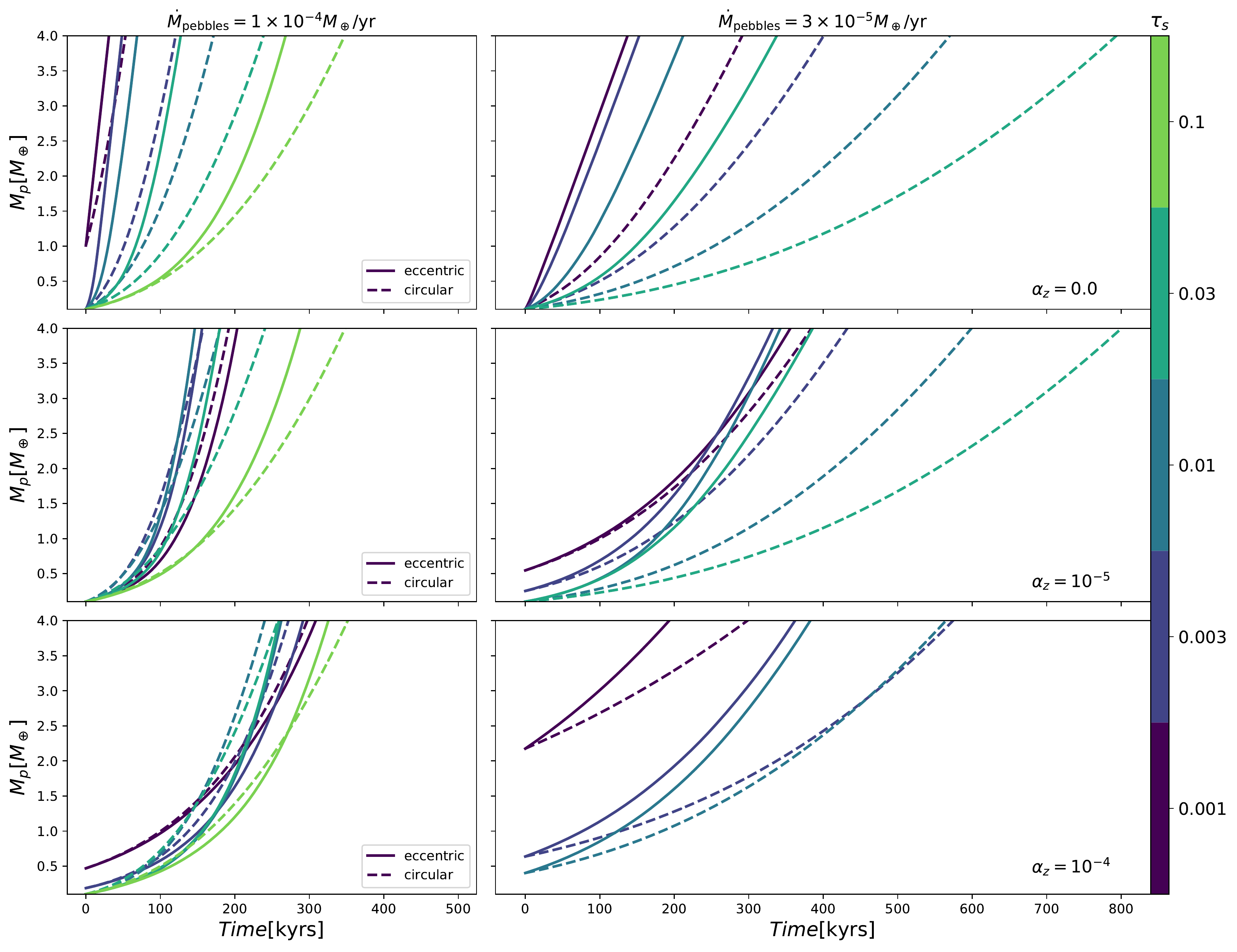}
    \caption{Planet mass as a function of time for the case $\eta=0.002$, for five different values of the dimensionless stopping time $\tau_s$, for a mass flux of pebbles of $10^{-4}\;M_\oplus.\textrm{yr}^{-1}$ (left) and $3\cdot 10^{-5}\;M_\oplus.\textrm{yr}^{-1}$ (right). In solid lines the results obtained for eccentric planets, in dashed lines the results for planets that remain on circular orbits. These results show that eccentricity driving can considerably lower the time needed to reach a given mass.}
    \label{fig:M_t}
\end{figure*}

In Fig.~\ref{fig:M_t} we present the mass history of the embryos for the disc models presented in \ref{fig:L_M}. The time integration is performed using the accretion rate interpolated for the instantaneous mass. We start with the minimum mass for which we have simulation date for each disc model, and present the evolution for both eccentric and circular cases from said mass. We observe a faster growth in the circular scenario only for $\alpha_z=10^{-4}$ and $\dot{M}_\text{pebbles}=10^{-4}\;M_\oplus.\textrm{yr}^{-1}$. In general, we observe that the fact that the planet becomes eccentric drastically shortens the time taken to reach the final mass, with respect to the case in which the planet would be restricted to a circular orbit.

\section{Discussion and conclusions}
\label{sec:disc-concl}
We have adapted our nested-meshes version of the \texttt{FARGO3D} code to perform simulations of a freely evolving, luminous planet in a thermally diffusive gaseous disc around a solar mass star. The implementation allowed us to improve the resolution of previous work \citep{2017arXiv170401931E}, and to get converged results for the eccentricity reached at larger time, or asymptotic eccentricity, by a low-mass planet as a function of its luminosity. We have performed an exploration of the parameter space to obtain the asymptotic eccentricity of a planet with semi-major axis $a=5.2$~au, as a function of the planet's mass and luminosity for the model of protoplanetary disc given in Tab.~\ref{tab:1}, which has values comparable to that of the \emph{Minimum Mass Solar Nebula} \citep[MMSN,][]{1981PThPS..70...35H}. When the planet's luminosity is in excess of $L_c$, the critical luminosity given by Eq.~\eqref{eq:Lc}, the asymptotic eccentricity is finite, slowly increases with the planet's luminosity, and tends to be independent of the planet mass for a given mass to luminosity ratio, when the planetary mass is smaller than the critical mass of Eq.~\eqref{eq:criticalmass}. For a given mass to luminosity ratio, higher mass planets tend to have smaller eccentricities. Finally, we find that the asymptotic eccentricity increases with the disc's aspect ratio, and scales with the latter for luminosities significantly larger than the critical luminosity.

In a second step, we have examined the feedback of the eccentricity on the accretion of pebbles, hence on the planet's luminosity. Assuming that the planet owes its luminosity exclusively to the accretion of pebbles, we have made use of the analytic expressions of \citet{2018A&A...615A.138L} and \citet{2018A&A...615A.178O} to obtain the accretion rate as a function of the pebbles' dimensionless stopping time, of the headwind velocity (or mismatch between the orbital velocity of the gas and Keplerian velocity) , the planet's eccentricity and the inward mass flux of pebbles. We have searched for the equilibrium eccentricity: the luminosity acquired by the planet for such eccentricity is precisely the luminosity that brings the planet to this eccentricity through thermal forces.

We find that the eccentricities reached by pebble accreting planets are comparable to the aspect ratio of the disc, which confirms earlier results \citep{2017A&A...606A.114C,2017MNRAS.465.3175M,2017arXiv170401931E}. For the disc parameters adopted, we find that planetary embryos below an Earth mass can reach eccentricities exceeding significantly the aspect ratio, when the dimensionless stopping time of the pebbles is smaller than $10^{-2}$. The eccentricity of planets more massive than the Earth depends weakly on their mass and on the size of the pebbles, and decays slowly with the mass flux of pebbles. The planets at the upper limit of the mass range considered in our study ($4\;M_\oplus$) still have important eccentricities, which suggests that the mechanism presented here could apply to significantly more massive planets and that it could be possible to detect observational signatures of this process in nearby protoplanetary discs. In none of our simulations have we found indefinitely growing eccentricities in the supersonic regime, as suggested by \citet{2017MNRAS.465.3175M}.

We mention hereafter a few shortcomings of our analysis.

\begin{enumerate}
\item \emph{We have assumed that the pebbles reach the core, and that their potential energy is entirely converted into heat, as implied by Eq.~\eqref{eq:15}.} The assumption that the pebbles reach the core without ablation is correct up to a fraction of an Earth mass \citep{2018A&A...611A..65B}. For cores more massive than half of an Earth mass, the pebbles are fully ablated before reaching the core \citep{2018A&A...611A..65B} and contribute to the formation of a high-$Z$ vapour close to the core \citep{2020A&A...634A..15B}. The luminosity provided by Eq.~\eqref{eq:15} is therefore an upper limit of the planet's luminosity. Only a detailed calculation, largely beyond the scope of this work, that takes into account the evolution of pebbles as they enter the envelope surrounding the embryo can provide an accurate value for the luminosity. We note however that the high-$Z$ vapour is located close to the core \citep{2020A&A...634A..15B} so that the energy ultimately released by the material out of which it is formed should not differ by a large factor from $GM_p/R_p$ (using the notation of Eq.~\eqref{eq:15}), minus the latent heat for vaporisation.
\item \emph{Planets with supersonic motion with respect to the gas can produce bow shocks which could destroy the pebbles before they can be accreted onto the planet} \citep{2018A&A...615A.138L}. When a planet has an eccentricity in excess of the aspect ratio, a fraction of its epicycle is described at supersonic velocity with respect to the underlying gas, and accretion may only occur in the subsonic part of the epicycle (near the apocentre and pericentre). When the eccentricity is larger than twice the aspect ratio, the planet is always supersonic with respect to the gas, and pebble accretion may be completely inhibited. This, in addition to the sharp decay of thermal forces in the supersonic regime, may provide a cut-off to the growth of eccentricity past the aspect ratio.
\item \emph{We have assumed the inclination to be negligible.} Should the planet be on an orbit with an inclination not small compared to the aspect ratio of the disc of pebbles, accretion would drop during the phases where its vertical excursion exceeds the thickness of the pebbles' disc. Note, however, that \citet{2017arXiv170401931E} have found that the growth of eccentricity is much faster than that of inclination, and that once the eccentricity plateaus to values comparable to the aspect ratio, the growth of the inclination is quenched. Starting from a nearly circular, coplanar orbit, a single embryo therefore reaches an essentially non-inclined, eccentric orbit.
\item \emph{We have neglected the contribution of planetesimal accretion to the planet's luminosity.} To give an idea of the magnitude of this effect, a constant mass accretion rate of planetesimals of $10^{-5}\;M_\oplus.\mathrm{yr}^{-1}$ would lead to a planetary luminosity $L\approx 2.9L_c(M/M_\oplus)^{-1/3}$ for the disc parameters of Tab.~\ref{tab:1} and an embryo density of $3$~g.cm$^{-3}$. Should large mass embryos owe most of their luminosity to planetesimal accretion \citep[e.g.][]{2021ApJ...915...62I}, a dedicated study of the accretion rate of planetesimals as a function of the planet's eccentricity should be undertaken. 
\item \emph{We have neglected the feed back of the heated region onto the accretion rate of pebbles}. The gas flow in the vicinity of the planet is modified by the energy released by the latter \citep{2017A&A...606A.146L, 2019A&A...626A.109C,2019MNRAS.482L.107P}, which in turns alters the drag force exerted on the solids. As long as this effect occurs deep inside the planet's Hill sphere, it should make no difference on the accretion rate of pebbles which are in the settling regime \citep{2017ASSL..445..197O}, whereas it might have some impact for those which are in the ballistic regime, which occurs for significantly eccentric embryos, past the peak of maximal accretion rate.
\end{enumerate}
From the results presented here it should be clear that eccentric, low-mass embryos (up to $\sim 1\;M_\oplus$) should be the norm rather than the exception. Whether this statement could be extended to larger masses depends on the ultimate value of the luminosity arising from pebble accretion, which requires a careful assessment, beyond the scope of this work, and from the additional luminosity coming from planetesimal accretion.

We finally mention that our analysis of section~\ref{sec:feed-back-eccentr} was done assuming a stationary, inward flow of pebbles, and a finite headwind velocity. If there are other embryos at larger radii, they may filter out a substantial fraction of the pebble mass flux, and the eccentricity achieved is lower than that predicted for a single embryo, whereas if one of the outer embryos exceeds the pebble isolation mass, it completely shuts off the pebble flow inside of its orbit, so that all the embryos interior to it become circular. The assumption of a stationary inward flow breaks down at pressure maxima, where the dust is trapped, and a specific treatment is warranted to investigate the behaviour of an embryo located in the vicinity of the trap \citep{2020A&A...638A...1M}. As the dust there remains available for accretion ``forever'', it is reasonable to expect that the accretion efficiency is much larger than in the bulk of the disc where pebbles that have drifted past the planet's Hill sphere (usually the majority of them) can no longer be accreted. As a consequence, low-mass planetary embryos trapped in dust rings should have significant eccentricities, and a growth time scale significantly shorter than if they were on circular orbits. Also, if these embryos grow preferentially by accretion of planetesimals, a treatment similar in spirit to that presented here is warranted, in which one should study the feedback of the embryos' eccentricity on the planetesimal accretion rate.

\section*{Acknowledgements}
The simulations performed for this work were executed on Piz Daint at the Swiss Supercomputing Center CSCS in Lugano and on the Vesta cluster of the University of Zurich. The authors gratefully acknowledge support from the Marcos Moshinsky Foundation through the Marcos Moshinsky Fellowship. F.M. gratefully acknowledges support from grants UNAM-DGAPA-PASPA and UNAM-DGAPA-PAPIIT IG-101-620, as well as the University of Nice-Sophia Antipolis and the Observatoire de la C\^ote d'Azur for hospitality.

\section*{Data availability statement}
The data underlying this article will be shared on reasonable request to the corresponding author.


\bibliographystyle{mnras}
\bibliography{biblio}

\appendix
\section{Validation of the thermal implementation}
\label{app:1}
\begin{figure}
    \centering
    \includegraphics[width=\columnwidth]{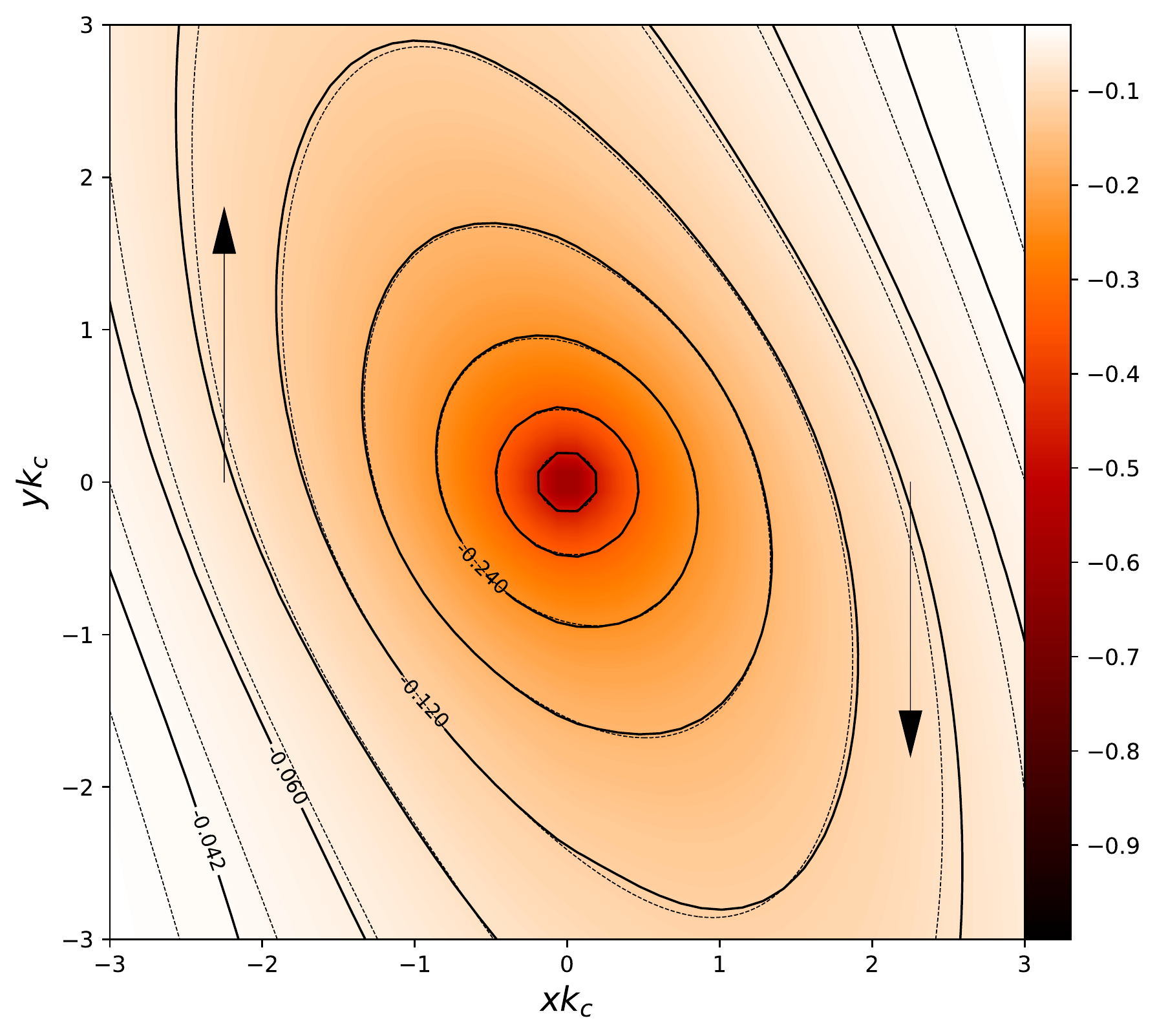}
    \caption{Perturbation of the surface density $\sigma^{'(0)}$ in units of $\gamma(\gamma-1)L/\chi c_s^2$ for $x^0_p=0$. In solid lines the numerical results with our implementation, in dashed lines the analytical results presented in Figure~1 of \citet{2017MNRAS.472.4204M}. We present the numerical results after $12$ orbits}
    \label{fig:sigma0}
\end{figure}

\begin{figure}
    \centering
    \includegraphics[width=\columnwidth]{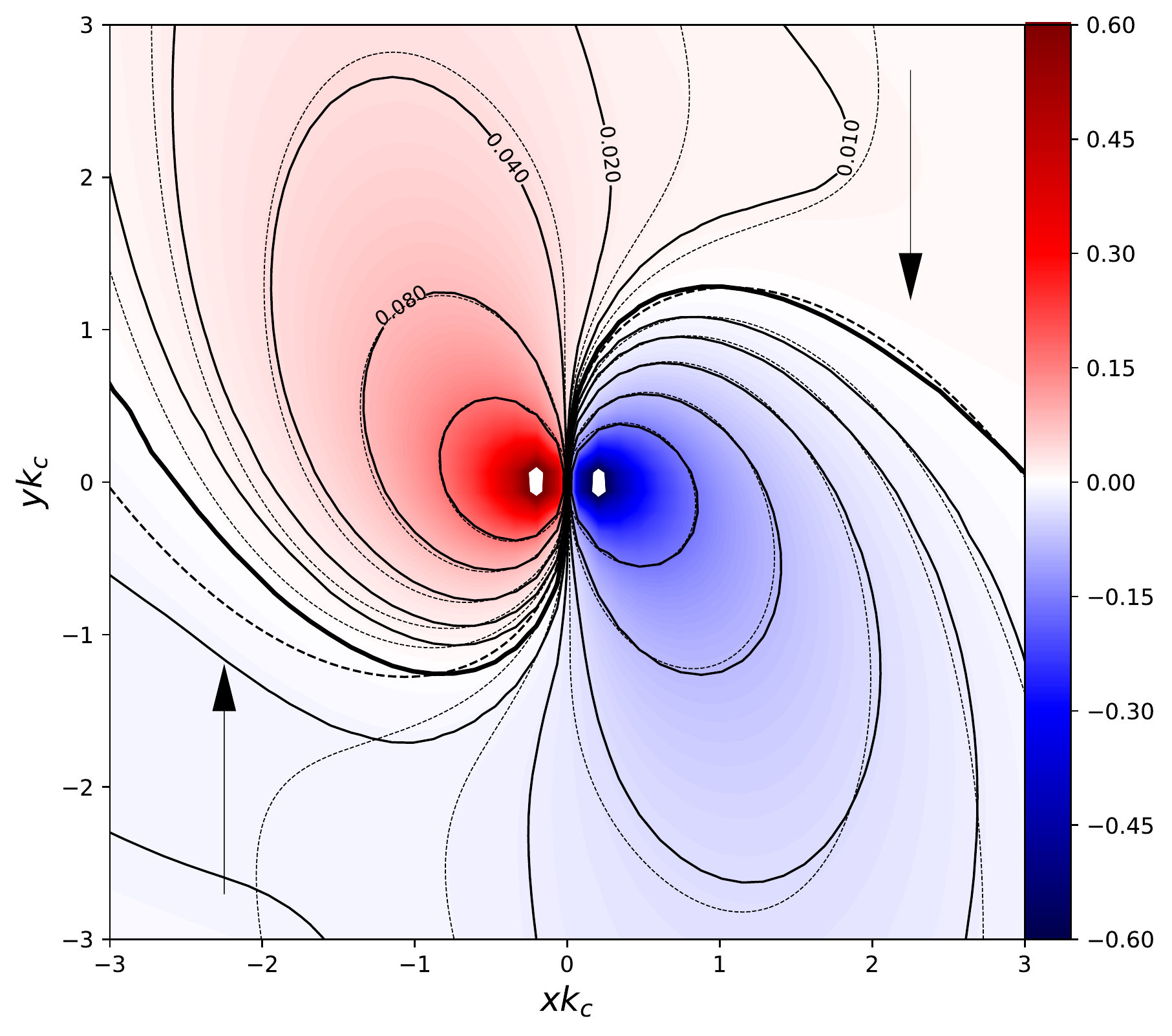}
    \caption{Map $\sigma^{'(1)}$ in units of $\gamma(\gamma-1)L k_c/\chi c_s^2$ obtained from the simulations with $x^0_p = \pm \Delta x$. In solid lines the numerical results with our implementation, in dashed lines the analytical results presented in Figure~1 of \citet{2017MNRAS.472.4204M}.}
    \label{fig:sigma1}
\end{figure}

In order to validate the numerical implementation of the disc's thermal diffusion and the planet's heat release, we performed three pairs of simulations with the purpose of reproducing the analytical results presented in Figure~1 of \citet{2017MNRAS.472.4204M}. The simulations were performed in the domain described by $\phi \in [-\pi/8,\pi/8]$, $r \in [0.9,1.1]$ and $\theta \in [\pi/2-0.06,\pi/2]$. The number of cells used was $(N_\phi,N_r,N_\theta) =(640,320,96)$. We used $\chi=4.5\times10^{-5}a^2\Omega_p$, a planetary mass $M_p=0.01M_c$ and, for the luminous cases, a luminosity $L=0.1L_c$. The first pair of simulations consists of a disc
with a surface density slope $\alpha=-1.5$ and a flaring index of $f=0.5$, which results in a null dimensionless coefficient $\eta = \frac{1}{3}(\alpha+2-f) = 0$, and therefore a vanishing planet's distance to corotation $x^0_p=\eta h^2 r_p = 0$. 
In Fig.~\ref{fig:sigma0} we present the perturbation of surface density $\sigma^{'(0)}$ in units of $\gamma(\gamma-1)L/\chi c_s^2$, obtained from the difference in density between two simulations, one performed with a luminous planet and another one performed with a non luminous planet. For the second and third pairs of simulations we chose $\alpha =-1.5 \pm 3\Delta x /(h^2 r_p)$, which results in a planet's distance to corotation of exactly one cell: $x^0_p = \pm \Delta x$. 
For each pair we subtract the map of surface density of the luminous and non-luminous cases, as previously, and by so doing obtain the maps $\sigma(x_p)$ and $\sigma(-x_p)$. We then construct the map $\sigma^{'(1)}=[\sigma(\Delta x)-\sigma(-\Delta x)]/(2\Delta x)$, where the maps $\sigma(\pm\Delta x)$ are shifted so that their corotations coincide. In Fig.~\ref{fig:sigma1}
we show $\sigma^{'(1)}$ in units of $\gamma(\gamma-1)L k_c/\chi c_s^2$. We present the results after $12$ orbits, which is roughly the diffusive time across the vertical extension of our domain ($a^2(\theta_\mathrm{max}-\theta_\mathrm{min})^2/\chi\approx 12.7$ orbits). After this time, the heat starts to accumulate in the mesh (at our lower boundary in colatitude there is a null temperature gradient) and the isocontours slowly expand with time. This effect has a negligible impact on the calculations presented here, as (i) the runs are performed on a shorter timescale and (ii) our eccentric planets move horizontally by an extent larger than $\lambda_c$.

\bsp	
\label{lastpage}
\end{document}